\documentclass[useAMS,usenatbib]{mn2e}
\usepackage{amssymb}
\usepackage{epsfig}

\def\mnras{MNRAS}  % Monthly Notices of the RAS
\def\apj{ApJ}      % The Astrophysical Journal
\def\apjl{ApJL}    % The Astrophysical Journal Letters
\def\apjs{ApJS}    % The Astrophysical Journal Supplement
\def\aap{A\&A}     % Astronomy and Astrophysics
\def\aaps{A\&AS}   % Astronomy and Astrophysics Supplement
\def\araa{ARA\&A}  % Annual Reviews of Astronomy and Astrophysics
\def\nat{Nature}   % Nature
\def\physrep{Physics Reports} % Physics Reports
\def\erg{\, \hbox{erg}}

\title[Heating and Cooling the ICM]{Towards a Holistic View of 
the Heating and Cooling of the Intracluster Medium}

\author[I. G. McCarthy et al.]{I. G. McCarthy$^{1}$\thanks{E-mail:
i.g.mccarthy@durham.ac.uk (IGM)}, A. Babul$^2$, R. G. Bower$^1$, 
M. L. Balogh$^3$
\\
\\
$^{1}$Department of Physics, University of Durham, South Road,
Durham, DH1 3LE\\
$^{2}$Department of Physics and Astronomy, University of Victoria,
Victoria, BC V8P 1A1, Canada\\
$^{3}$Department of Physics and Astronomy, University of Waterloo,
Waterloo, ON, N2L 3G1, Canada}

\begin{document}

\date{Accepted XXXX. Received XXXX; in original form XXXX}

\pagerange{\pageref{firstpage}--\pageref{lastpage}} \pubyear{2006}

\maketitle

\label{firstpage}

\begin{abstract}

X-ray clusters are conventionally divided into two classes:
``cool core'' (CC) clusters and ``non-cool core'' (NCC) clusters.
Yet relatively little attention has been given to the origins of 
this apparent dichotomy and, in particular, to the 
energetics and thermal histories of the two classes.  We develop 
a model for the entropy profiles of clusters starting from the 
configuration established by gravitational shock heating and 
radiative cooling.  At large radii, gravitational heating 
accounts for the observed profiles and their scalings well.  
However, at small and intermediate radii, radiative cooling and 
gravitational heating cannot be combined to explain the observed 
profiles of either CC or NCC clusters.  The inferred entropy 
profiles of NCC clusters require that material is ``preheated'' 
prior to cluster collapse in order to explain the absence of low 
entropy (cool) material in these systems.  We show that a 
similar modification is also required in CC clusters in order to 
match their entropy profiles at intermediate radii.  In CC 
clusters, this modification is unstable, and an additional 
process is required to prevent cooling below a temperature of a 
few keV.  We show that this can be achieved by adding a 
self-consistent AGN feedback loop in which the lowest-entropy, 
most rapidly cooling material is heated and rises buoyantly 
to mix with material at larger radii.  The 
resulting model does not require fine tuning and is in excellent 
agreement with a wide variety of observational data from {\it 
Chandra} and {\it XMM-Newton}, including entropy and gas density
profiles, the luminosity-temperature relation, and high 
resolution spectra.  The spread in cluster core morphologies is 
seen to arise because of the steep dependence of the central
cooling time on the initial level of preheating. Some of the
other implications of this model are briefly discussed.

\end{abstract}


\section{Introduction}

High quality X-ray data from the {\it Chandra} and {\it 
XMM-Newton} X-ray telescopes have spurred a great deal of 
excitement and debate over the competition between radiative 
cooling and non-gravitational heating in galaxy clusters.  In
particular, the association of X-ray cavities (``bubbles'')
with diffuse radio emission in many nearby ``cool 
core'' (CC) clusters (e.g., Fabian et al.\ 2000; David et al.\ 
2001; B{\^i}rzan et al.\ 2004), and the lack of X-ray spectral 
signatures of cool ($< 2$ keV 
or so) gas in such systems (e.g., Peterson et al.\ 2003) demonstrate
that heating processes may be at least as important as cooling.
Furthermore, we have detailed confirmation that the X-ray 
scaling properties of groups and clusters have been significantly 
modified by both radiative cooling and non-gravitational heating 
(e.g., Voit et al.\ 2002; McCarthy et al.\ 2004 (M04); Kay et 
al.\ 2004). As a consequence, greater  theoretical attention is 
being given to possible heating mechanisms, such as thermal 
conduction (e.g., Narayan \&  Medvedev 2001) and heating due to 
active galactic nuclei (AGN) and their associated bubbles (e.g., 
Binney \& Tabor 1995; Churazov et al.\ 2001; Mathews et al.\ 2004; 
Dalla Vecchia et al.\ 2004; Roychowdhury et al.\ 2004; Cattaneo 
\& Teyssier 2007).

While a great deal has been learnt about the impact of heating 
and cooling on the intracluster medium (ICM) in recent years 
(see Voit 2005 for a recent review), we still have not obtained 
a complete picture and many problems remain unresolved. 
One important issue that has received very little attention
is that the energy required to maintain a cluster in a given 
state can be very different from the energy required to set it in 
a particular configuration in the first place.  Thus, while 
comparing the {\it present-day} heating and cooling rates of the 
ICM is clearly important, by its nature this test does not 
address the question of how the ICM reached its current 
configuration to begin with.

Another important issue that is related to the above and 
remains relatively unexplored is the apparent dichotomy 
between ``cool core'' (CC) and ``non-cool core'' (NCC) clusters.  
Typically, a cluster is classified as being a CC (NCC) 
cluster based on the presence (absence) of a central positive
temperature gradient\footnote{The presence of a central 
positive temperature gradient is taken as an indicator 
of the 
importance of radiative cooling in the cluster core.  Indeed, 
massive clusters with temperature dips typically have relatively 
short central cooling times in comparison to those clusters 
without positive temperature gradients.  However, the observed 
distributions of central temperature, central cooling time, and 
central gas density (see, e.g., Fig.\ 7) appear to be more or 
less continuous across the cluster population, and so the exact 
dividing line used to assign CC/NCC status is somewhat arbitrary.  
This is not of concern for the present study, since we seek to 
explain the {\it entire} cluster population with the same 
physical model (see \S 4-5).}.  If defined in this way, one 
notes that the vast majority of published studies 
based on high quality {\it Chandra} and {\it XMM-Newton} data 
have been focused on the subset of CC clusters.  However, 
studies based on data from the previous generation of X-ray 
satellites suggest that these systems could represent less than 
half the nearby cluster population, and perhaps an even smaller 
fraction of all systems at high redshifts (see \S 3).  What is 
the origin of the NCC systems,
and what is their relation to the CC clusters?  The prevailing 
view, based on anecdotal evidence from the previous generation 
of satellites, is that they are the result of massive cluster 
mergers.  However, there is now a growing body of theoretical 
work based on semi-analytic methods, idealised simulations, and 
full cosmological simulations (e.g., M04; Motl et al.\ 2004; 
Borgani et al.\ 2004; O'Hara et al.\ 2005; Balogh et al.\ 2006; 
Poole et al.\ 2006; 2007) that suggests mergers by themselves 
cannot disrupt cool cores for significant periods of time and 
are not responsible for the large spread in cluster core 
morphologies.  The virtual 
absence of systems that resemble observed NCC systems (in terms 
of their gas radial profiles and stellar mass fractions) in 
cosmological simulations that include the effects of radiative 
cooling is perhaps the most compelling argument that mergers are 
not the answer.  Some form of extreme non-gravitational heating 
appears to be required to explain these systems.  One 
possibility, suggested originally by M04, is that the origin of 
NCC systems is linked to an early episode of entropy injection 
(i.e., ``preheating'').

An obvious question is whether preheating might also shape the 
properties of CC clusters.  At first sight this seems unlikely.  
The short central cooling times of observed CC clusters imply 
that without sufficient {\it present-day} heating, radiative 
cooling would rapidly establish extreme centrally-peaked gas 
density profiles and sharply declining temperature profiles with 
$T \rightarrow 0$ as $r \rightarrow 0$.  However, recent 
observations have convincingly demonstrated that present-day CC 
clusters are not in the state just described --- their central 
cores are warmer than expected (e.g., Peterson et al.\ 2003).  
On the other hand, the observed power output from AGN activity 
appears to be only just sufficient to counteract radiative 
cooling losses (e.g., Dunn \& Fabian 2006) in the clusters' 
present-day configuration.  If the central 
cores were previously cooler and denser than at present (as 
expected prior to the first heating episode), energy deposition 
of $\sim 10^{62}$ ergs would be required to drive the 
systems to their present configuration (see \S 4).  This is larger 
than that provided by even the most extreme heating events 
observed to date.  An alternative is that CC 
systems could have evolved to their present states from a 
hotter earlier configuration --- perhaps from a state similar to 
the NCC clusters we observe today.  In this paper, we will show 
evidence that today's CC clusters are part of a continuous 
distribution of preheating levels. The apparent dichotomy arises 
from the steep dependence of the central cooling rate on the 
initial level of heating.

In order to begin to address these important issues, the role of 
non-gravitational heating in clusters needs to be quantified.
Fortunately, progress can now be made on this front.  In 
particular, cosmological simulations can now robustly predict 
the properties of the ICM in the non-radiative limit (e.g., 
Frenk et al.\ 1999).  For example, Voit et al.\ (2005) have 
recently shown that the entropy structure of simulated clusters 
is independent of cluster mass (more precisely, the entropy 
profiles are self-similar) and is robust to simulation technique.  
Obviously, within the cooling radius the effects of radiative 
cooling are important and need to be factored in.  This is 
relatively straightforward to accomplish with the aid of 
idealised 1-D hydrodynamic simulations (e.g., Kaiser \& Binney 
2003; Oh \& Benson 2003; M04; Hoeft \& Br{\"u}ggen 2004).  We 
refer to the model with the effects of cooling factored in as the 
``pure cooling model''.

The strategy we adopt in this paper is to use the difference
between the pure cooling model and the observed profiles of
clusters in order to quantify the non-gravitational contribution
to the heating in both CC and NCC systems.  Based on this
empirical measurement we then explore various models for the
heating mechanism.   We demonstrate that a model that 
combines an early bout of preheating with ongoing 
AGN heating at the centre (that operates should the central 
cooling 
time of the gas following preheating be relatively short) 
provides a full description of the observed range of cluster 
properties.

The present paper is organised as follows.  In \S 2, we present 
a description of the cluster models, including the pure cooling 
model.  In \S 3, we compare the pure cooling model to a wide 
range of observations of CC 
clusters and to available data of NCC clusters.  We calculate 
the (non-gravitational) heating energy required to explain the 
differences between the observed systems and the pure cooling 
model in \S 4.  In \S 5, we propose a physical heating model 
that adheres to these requirements and provides an excellent 
match to the observational data.  Finally, in \S 6, we present a 
summary of our findings and some conclusions.  We 
assume the following cosmological parameters 
throughout: $h = 0.7$, $\Omega_m = 0.3$, $\Omega_\Lambda = 0.7$, 
and $\Omega_b = 0.02 h^{-2}$.

\section{Cluster Models}

\subsection{The baseline gravitational model}

The setup of our baseline cluster models is inspired by the 
results of recent cosmological simulations.  The dark matter is 
assumed to follow a NFW distribution (Navarro et al.\ 1997):

\begin{equation}
\rho_{\rm dm}(r) = \frac{\rho_s}{(r/r_s)(1+r/r_s)^2}
\end{equation}

\noindent where $\rho_s = M_s/(4 \pi r_s^3)$ and

\begin{equation}
M_s=\frac{M_{200}}{\ln(1+r_{200}/r_s)-(r_{200}/r_s)/(1+r_{200}/r_s)}
\end{equation}

In the above, $r_{200}$ is the radius within which the mean 
density is 200 times the critical density, $\rho_{\rm crit}(z)$, 
and $M_{200} \equiv M(r_{200}) = (4/3) \pi r_{200}^3 \times 200 
\rho_{\rm crit}(z)$.

To fully specify the dark matter density profile of a halo of mass 
$M_{200}$, a value for the scale radius, $r_s$, must be selected.  
The scale radius can be expressed in terms of the concentration 
parameter, $c_{200} \equiv r_{200}/r_s$.  We adopt the 
mass-concentration ($M_{200}$-$c_{200}$) relation recently 
derived from the {\it Millennium Simulation} (Springel et al. 
2005) by Neto et al.\ (2007). For example, for a 
massive cluster with $M_{200} = 10^{15} M_\odot$, Fausti Neto et 
al.\ find $c_{200} \approx 4.3$ (see also Eke et al.\ 2001).  

For the gaseous baryonic component, we adopt the baseline 
entropy profile of Voit et al.\ (2005), where the entropy $S$ is 
defined\footnote{Strictly speaking, $S$ is not the 
thermodynamic specific 
entropy ($s$) but rather, a proxy for the latter that has gained 
currency in X-ray astronomy because it can be easily constructed 
out of observed quantities $n_e$ and $T$.   Our ``entropy proxy'' 
and the thermodynamic specific entropy are, however, related to 
each other as:  $ds \propto d\ln{S}$ (see Balogh et al. 1999).
For the sake of clarity, we keep to current practise and refer to 
$S$ as ``entropy'' throughout this paper.} as $k_B T / 
n_e^{2/3}$.  Briefly, Voit et al.\ found  that the entropy 
profiles of clusters formed in non-radiative cosmological 
simulations are nearly self-similar (i.e., do not depend on 
cluster mass).  The form of this profile at large radii is 
robust to simulation technique, with clusters simulated using 
either a Lagrangian smoothed particle hydrodynamics (SPH) code 
or a Eulerian adaptive mesh refinement (AMR) code having nearly 
the same entropy structure.  Accordingly, we adopt the following 
entropy profile at large radii, which is consistent with both 
their SPH and AMR results (see their equations 9 and 10):

\begin{equation}
\frac{S(r)}{S_{200}} = 1.47 \biggl(\frac{r}{r_{200}} 
\biggr)^{1.22}
\end{equation}

The characteristic entropy of the cluster, $S_{200}$, is defined 
as:

\begin{eqnarray}
S_{200} & \equiv & \frac{k_B T_{200}}{[n_{e,200}]^{2/3}} \\
& = & \frac{G M_{200} \mu m_p}{2 r_{200} [200 f_b \rho_{crit}(z) 
/ (\mu_e m_p)]^{2/3}} 
\nonumber \\
& \approx & 2561 \ {\rm keV \ cm}^2 
\ \biggl(\frac{M_{200}}{10^{15} M_\odot} 
\biggr)^{2/3} \biggl(\frac{h}{0.7} \ \frac{f_b}{0.13} 
\biggr)^{-2/3}
\nonumber
\end{eqnarray}

In the above, $f_b \equiv \Omega_b/\Omega_m$ is the universal 
mass ratio of baryonic to total matter.  We adopt $\mu = 0.59$ 
and $\mu_e = 1.14$ (where $\mu$ and $\mu_e$ are the mean 
molecular weight and the mean molecular weight per free 
electron, respectively), appropriate for a gas with metallicity of 
$Z = 0.3 Z_\odot$.  The last entry in eqn. (4) assumes $z = 0$.

At small radii, cores are present in the entropy profiles of 
both the SPH and AMR simulated clusters.  The origin of this core 
is at least partially attributable to energy exchange between the 
gas and the dark matter (e.g., Lin et al.\ 2006; McCarthy et 
al.\ 2007a).  However, the SPH and AMR codes predict 
significantly different core amplitudes (see also Frenk et al.\ 
1999).  The origin of this discrepancy is presently unclear.  
Fortunately, it is inconsequential for our purposes since the 
core is typically confined to radii where the cooling 
time\footnote{The cooling time, $t_{\rm cool}$, is defined as 
$[3 n k_B T] / [2 n_e n_H \Lambda(T)]$.} of the gas is short; 
i.e., the core is smaller than the cooling radius.  Radiative 
cooling always modifies the entropy of the gas within the 
cooling radius, driving $S \rightarrow 0$ as $r \rightarrow 0$.  
Thus, the initial conditions of the gas at small radii have 
little or no bearing on the properties of the (pure cooling) 
model clusters.  But for specificity, we adopt a core amplitude 
of $0.06 S_{200}$, which is consistent with the SPH simulations 
of Voit et al.\ (2005).

With the above dark matter density and gas entropy profiles, the 
remaining properties of the gas (e.g., temperature and density) 
are determined by assuming the gas is in hydrostatic equilibrium 
(hereafter, HSE) within the dark matter-dominated 
gravitational potential well:

\begin{equation}
\frac{dP(r)}{dr} = - \frac{G M_{\rm dm}(r)}{r^2} \rho_{\rm 
gas}(r) 
\end{equation}

\noindent where for simplicity we have neglected the self-gravity 
of the gas.  

A boundary condition is required in order to solve equation (5).  
Non-radiative cosmological simulations of clusters demonstrate that 
the ratio of gas to total mass converges to nearly the universal 
ratio at large radii (e.g., Frenk et al.\ 1999; Kay et al.\ 
2004; Crain et al. 2007).  Accordingly, we iterate $P(r_{200})$, 
the gas pressure at the outer boundary of the clusters, until 
the following condition is satisfied:

\begin{equation}
\frac{M_{\rm gas}(r_{200})}{M_{200}} = b f_b
\end{equation}

The bias factor, $b$, is set to $0.9$ for our models (see Crain 
et al.\ 2007).

The above fully specifies the properties of the model clusters 
prior to modification of the gas owing to the effects of 
radiative cooling.

\subsection{The effects of radiative cooling}

\subsubsection{The pure cooling model}

To compute the effects of radiative cooling, we use the 
time-dependent 1-D hydro algorithm of M04 (see also Kaiser \& 
Binney 2003; Oh \& Benson 2003).  We give a brief description of 
the model here, but refer the reader to \S2.2 of that study for 
additional details.

The baseline profiles derived in \S 2.1 are discretised onto a 
non-uniform Lagrangian grid (i.e., the various quantities, such 
as temperature and density, are interpolated onto a grid of gas 
mass shells).  A non-uniform grid is adopted so as to finely 
resolve the central regions of the cluster without investing 
a large computational effort on the outermost regions, where 
the effects of cooling are minimal.

We evolve the system with time by following the effects of 
radiative cooling (and, later, non-gravitational heating) on the 
gas entropy.  Entropy is the most physically interesting quantity 
to track when cooling and non-gravitational heating are relevant, 
since convection will order the distribution such that the 
lowest entropy gas is at the bottom of the cluster potential 
well and the highest is located at the cluster 
periphery\footnote{But note convection plays a role even in the 
absence of radiative cooling and/or non-gravitational heating.  
The entropy of the vast bulk of the ICM is largely generated via 
gravitational shock heating during mergers.  Since not all 
regions of a cluster are shock heated to the same degree during 
a particular merger, some convective mixing results (McCarthy et 
al.\ 2007a).}.  The temperature and density profiles of the ICM 
can therefore really just be thought of as manifestations of the 
underlying equilibrium entropy configuration (e.g., Voit et al.\ 
2002; 2003; Kaiser \& Binney 2003).  Furthermore, entropy is 
more straightforwardly linked to cooling and non-gravitational
heating since cooling always decreases the entropy and heating
always raises it.  Note this is not the case with the density or
temperature, since heating the gas can cause it to expand and,
therefore, potentially {\it lower} both quantities, as is the
case in a self-gravitating gas cloud.

Radiative cooling reduces the entropy of a parcel (shell) 
of gas according to

\begin{equation}
\frac{d\ln{S^{3/2}}}{dt} = - \frac{\mu m_p n_H n_e 
\Lambda(T)}{\rho_{\rm gas} k_B T} ,
\end{equation}

\noindent where $\Lambda(T)$ is the bolometric cooling function, 
which we calculate using a Raymond-Smith plasma model (Raymond \& 
Smith\ 1977) with a metallicity fixed to $Z = 0.3 Z_\odot$. 

Equation (7) is integrated over a time interval $dt$.  The 
size of this time step is computed by adhering to the standard 
Courant condition: $dt \leq \Delta r/c_s$, where $\Delta r$ is the 
gas mass shell size and $c_s$ is its speed of sound.  We 
calculate this quantity for all gas mass shells, find the shell 
with the minimum time step, and apply this time step uniformly 
to all shells.  

Integrating equation (7) yields a new (lower) value for the 
entropy of each gas mass shell.  If we assume that the total 
gravitational potential of the cluster, which is dominated by 
dark matter, does not evolve with time, we can use this new 
$S(M_{\rm gas})$ distribution\footnote{The quantity $S(M_{\rm 
gas})$ is most straightforwardly thought of in terms of its 
inverse, $M_{\rm gas}(S)$, which is the total mass of gas with 
entropy lower than $S$.} to update the other gas variables (e.g., 
temperature and density) as a function of time by simultaneously 
solving the equations of HSE and mass continuity (see M04).

We carry the integration forward in time until the onset of 
catastrophic cooling (i.e., when $T$ or $S$ of the innermost 
mass shell drops to zero).  It is at this point in time that a 
central AGN is assumed to ``switch on'' and heat the gas (e.g., 
Kaiser \& Binney 2003).  We refer to the physical conditions of 
the ICM at the onset of catastrophic cooling, just prior to AGN 
switching on, as the ``pure cooling model''.

Strictly speaking, a central AGN could switch on before the 
onset of catastrophic cooling since a massive black hole sitting 
at the centre of the cluster acts as a mass sink.  According to 
the spherically symmetric model of Bondi (1952), the accretion 
rate, $\dot{M}_{\rm bh}$, onto a central black hole is given by

\begin{eqnarray}
\dot{M}_{\rm bh} & \approx & 1.3\times10^{-5}
M_\odot \ {\rm yr}^{-1} \nonumber \\
& & \times \ \ \biggl(\frac{M_{\rm
bh}}{10^9 M_\odot}\biggr)^2 \biggl(\frac{S}{100 {\rm \ keV
\ cm}^2}\biggr)^{-3/2} \nonumber
\end{eqnarray}

\noindent where $M_{\rm bh}$ is the mass of the black hole and
$S$ is the entropy of the gas being accreted (e.g., Babul et
al.\ 2002).  The heating rate of the ICM, $L_{\rm bh}$, is 
typically assumed to be a fraction of the rest mass energy of the
accreting material, i.e., $L_{\rm bh} = \epsilon \dot{M}_{\rm bh}
c^2$, where $\epsilon$ is the efficiency factor. Taking typical
values for $M_{\rm bh}$ and $\epsilon$ (we assume $10^9 M_\odot$
and $0.1$, respectively), we find that in order to offset 
radiative cooling at a rate of $\sim 10^{45}$ ergs s$^{-1}$ within 
$r_{\rm cool}$, $S$ must be lower than $1$ keV cm$^2$.  This 
is much lower than the minimum cluster entropy of $\sim 150$ keV 
cm$^2$ for a $M_{200} = 10^{15} M_\odot$ cluster found in {\it 
non-radiative} cosmological simulations (see \S 2.1).  If we 
evolve this baseline gravitational model with radiative
cooling until the central entropy reaches $S \approx 1$ keV
cm$^2$, we find that the resulting gas profiles are nearly
identical to those of the pure cooling model.

It is important to note that there are essentially no free
parameters in the pure cooling model.  It is the configuration
that must result when radiative cooling is applied to the
baseline gravitational model specified in \S 2.1.  The
gravitational baseline model itself is fixed entirely by the
total cluster mass $M_{200}$ and the baryon fraction $f_b$,
which is set by the cosmological ratio  $\Omega_b/\Omega_m$.

The model we are using is evidently idealised.  In Appendix A, 
we briefly discuss the approximations that have been made and 
their validity.  We show that the model provides an accurate 
description of the cooling problem.

Finally, we re-emphasise that we are not advocating the pure 
cooling model as a legitimate physical model for the ICM.  The 
sole purpose of the model is to constrain the energetics (of 
non-gravitational heating) required to drive the observed 
systems to their present configurations.  This is done 
by first empirically modifying the entropy profile of the pure 
cooling model as a function of radius so that it matches the 
observed systems and then comparing the total energies of the 
modified and pure cooling distributions.  Quantifying the 
energetics in turn allows us to constrain physical heating 
models for the ICM, which we explore in \S 4-5.

\section{Comparison with Observations}

To obtain an unbiased picture of non-gravitational heating in 
clusters one would ideally like to compare the models to a 
large, statistically-representative sample observed at high 
spatial and spectral resolution.  While {\it Chandra} and {\it 
XMM-Newton} data is of sufficiently high quality to achieve this 
goal, the construction of large representative samples of 
uniformly-analysed clusters observed with these telescopes has 
been slow going.  Much of the early interest has been focused on 
a subset of the most X-ray luminous systems in the universe; 
i.e., so-called ``cool core'' (hereafter, CC) systems.  By 
definition, these are systems that are characterised by 
their central positive temperature gradients and short 
central cooling times, with $t_{\rm cool} < 3-4$ gigayears (see, 
e.g., Sanderson et al.\ 2006; Dunn \& Fabian 2006).  

However, it should be reiterated that CC systems are a particular 
subset of the cluster population.  For example, Peres et al.\ 
(1998) found that approximately 50\% of clusters in the 
flux-limited sample of Edge et al.\ (1990) have central 
cooling times that {\it exceed} 4 gigayears. A similar result 
has recently been reported by Chen et al.\ (2007) for the 
HIFLUGCS flux-limited sample (also based on {\it ROSAT} and {\it 
ASCA} data).  Moreover, because of their high luminosities, 
one should expect flux-limited samples (as opposed to ideal
volume- and mass-limited samples) to be biased in favour of 
including CC systems.  This suggests that NCC systems could 
actually represent more than half the nearby cluster population, 
and perhaps an even larger fraction of more distant systems 
(Vikhlinin et al.\ 2007).

In light of the above, we have split the comparison to 
observations of CC and NCC clusters into two sections.  In \S 
3.1, we compare the models to new {\it Chandra} and {\it 
XMM-Newton} data of CC clusters.  In \S 3.2, we compare the 
models to {\it ROSAT} and {\it ASCA} data of NCC clusters.

We attempt to treat the comparison between the observational data
and the models in a fair fashion.  First, where appropriate the 
observed quantities are adjusted to account for any differences 
in the assumed cosmologies.  For the models, we compute 
``spectral'' temperatures, as opposed to emission-weighted 
temperatures (see, e.g., Mazzotta et al.\ 2004; Rasia et al.\ 
2005; Poole et al.\ 2007), by first generating synthetic spectra 
using the MEKAL 
plasma model (Mewe et al.\ 1985; Kaastra 1992; Liedahl et al.\ 
1995) with a metallicity fixed to $Z = 0.3 Z_\odot$ and a typical 
Galactic absorption column density of $N_H = 3 \times 10^{20}$ 
cm$^2$ and then fitting the spectra using a single-temperature 
MEKAL model over the passband of the relevant X-ray instrument.  
Finally, for the sake of consistency, the same scaling relations 
used by the observers to estimate, e.g., characteristic radii 
such as $r_{500}$ are also applied to the models.

We restrict the comparison to high mass clusters only, 
with mean spectral temperatures of 3 keV or greater.  The reason 
for this is as follows.  As data quality decreases with cluster 
mass, 
owing to the decreasing X-ray luminosity with mass, it becomes 
more and more difficult to reliably measure, for example, 
spatially-resolved temperature and entropy profiles.  As a 
result, there is an increasing need to take into account more 
carefully the details of the observing conditions by, e.g., 
folding the instrumental response of the relevant X-ray 
instrument into the theoretical models and
then reducing the data in the same fashion as the observers.
This is beyond the scope of the present study.  However, recent
analyses of mock {\it Chandra} observations of simulated massive 
clusters have demonstrated that, for example, the standard 
deprojection techniques for deriving the temperature and density 
profiles of the ICM from X-ray data should be accurate to within 
a few percent under typical observing conditions.  This is good 
news for theorists since it means a detailed accounting of the 
observing conditions is typically not required for massive 
(high temperature) clusters such as those studied below (e.g., 
Rasia et al.\ 2005; Nagai et al.\ 2007).

\subsection{Cool core (CC) clusters}

In this section we show that a simple, empirical modification to 
the pure cooling entropy profile matches observed entropy 
profiles of CC clusters remarkably well.  This modification also 
{\it simultaneously} brings the predicted density profiles, 
luminosity-temperature relation, and X-ray spectra into 
excellent agreement with observations. The simultaneous 
agreement with all these observations is non-trivial, and 
demonstrates that a wide range of observable 
quantities can be explained by a single modification to the 
entropy structure.

\subsubsection{Physical Entropy Profiles}

In Figure 1, we plot three different sets of observed entropy 
profiles.  Donahue et al.\ (2006) (hereafter, D06) and Vikhlinin 
et al. (2006) (hereafter, V06) used {\it Chandra} data to derive 
the ICM radial profiles for two different samples of CC systems.  
In both studies the profiles were fitted with smooth parametric 
models.  We use their models and associated best-fit parameters 
to reconstruct the entropy profiles of their systems (see Table 
6 of D06 and Tables 1-3 of V06).  The 
third data set is from Pratt et al.\ (2006) (hereafter, P06), who 
derived the ICM radial profiles of another sample of CC clusters 
using {\it XMM-Newton} data.  In all three studies, the 
ICM radial profiles were derived using a standard deprojection 
technique incorporating the spatially-resolved projected 
temperature and surface brightness profiles of their clusters.

The sample of D06 was constructed from archival {\it 
Chandra} data, typically limiting their analysis to relatively 
small radii ($r < 300$ kpc).  The observations of V06 and P06, 
by contrast, were designed to probe out to much larger radii ($r 
\sim 1$ Mpc).  The three data sets therefore complement each 
other nicely, providing good continuous radial coverage out to 
$\sim r_{500}$.  The observed entropy profiles show a high 
degree of uniformity, particularly at large radii.  This is 
in spite of the fact that the mean temperatures of their systems 
span more than a factor of 3 or so.  Reassuringly, the entropy 
profiles of the three studies show very similar properties at 
intermediate radii ($30$ kpc $< r < 250$ kpc), where there is 
significant overlap between them.  At very small radii, a 
minimum entropy ranging from $S \approx 5-15$ keV cm$^2$ is 
apparent in the data of D06, while at large radii the profiles 
appear to asymptote to a near powerlaw distribution.

\begin{figure}
\centering
\includegraphics[width=8.4cm]{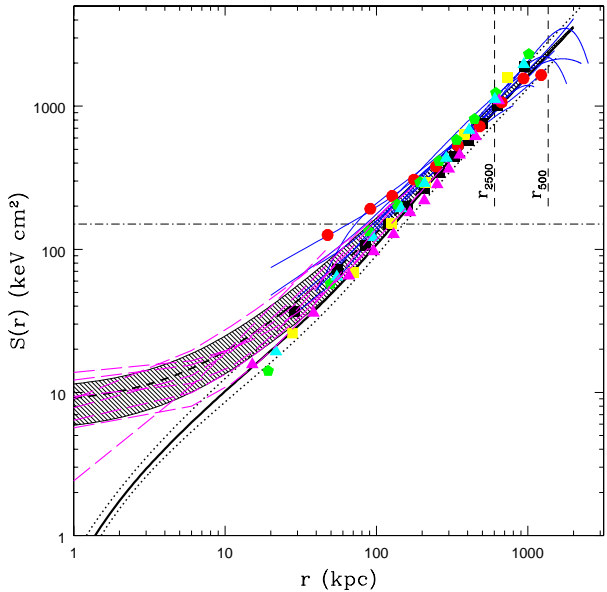}
\caption{Comparison of the {\it Chandra} entropy profiles of D06
and V06 and the {\it XMM-Newton} entropy profiles of P06 with
the predictions of the pure cooling model.  The thin dashed
magenta and solid blue curves represent fits to the entropy
profiles of D06 and V06, respectively, while the points
represent
the data of P06 (kindly provided by G. Pratt).  The solid
black curve represents the predictions of the pure cooling model
for a cluster with a typical total mass $M_{200} = 10^{15}
M_{\odot}$, while the two dotted curves show the predictions for
systems with $M_{200} = 5\times10^{14} M_\odot$ and
$2\times10^{15} M_\odot$ (bottom to top).  The shaded region
scatter) and short-dashed curve (median) represent the linear
modification to the pure cooling entropy profile (see eqn. 8)
for a cluster with $M_{200} = 10^{15} M_\odot$.  The horizontal
dot-dashed line indicates the typical entropy core amplitude for
a $M_{200} = 10^{15} M_\odot$ in the {\it non-radiative} SPH
simulations of Voit et al.\ (2005).
 }
\end{figure}

The entropy profiles predicted by the pure cooling model 
for three different masses are also plotted in Fig.\ 1.  This 
mass range is chosen so that the corresponding range in mean 
spectral temperatures is comparable to that of the observed 
clusters.  A comparison of the observed profiles to the profiles 
predicted by the pure cooling model leads us to conclude that: 
(1) at small radii ($r < 300$ kpc or so), the observed systems 
show 
evidence for excess entropy relative to the pure cooling model; 
and (2) at larger radii, where the effects of radiative cooling 
are minimal, the observed systems tend to converge to the 
theoretical distributions.  We comment on this second point in 
more detail below.

Voit \& Donahue (2005) recently reached the same conclusions by 
comparing a similar cooling-only model (see Voit et al.\ 
2002) with the profiles of D06.  (However, since the data of D06 
is limited to $r < 300$ kpc, the second conclusion was less 
evident.)  Like Voit \& Donahue (2005), we find that modifying 
the pure cooling model by adding an entropy pedestal (i.e., a 
constant value to all radii) of $S_0 \sim 10$ keV cm$^2$ 
results in reasonable agreement with the observed profiles; 
however at intermediate radii of roughly $30$ kpc $< r <$ 300 kpc 
a pedestal of this amplitude produces entropies that are 
somewhat too low compared to the observed profiles.  We find that 
the following linear relation

\begin{equation}
S(r) = 1.9^{+0.9}_{-0.7} \ \biggl(\frac{r}{1 \ \ h^{-1} \ {\rm 
kpc}} \biggr) + 8.0^{+2.0}_{-3.0} \ \ {\rm keV \ cm}^2
\end{equation}

\noindent yields a better fit to data at radii of $r <
300$ kpc or so, beyond which the data map closely onto the 
baseline profile.  We therefore construct a ``linear 
modification'' in which the entropy profile follows equation (8)
until it intersects\footnote{Note that the radius at which these
profiles intersect varies with total cluster mass.} with the pure
cooling entropy profile.  However, it is worth bearing in mind 
that the clusters studied by D06, V06, and P06 all have implied 
central cooling times of less than 1 gigayear or so and 
therefore represent somewhat extreme examples of CC clusters 
(what would previously been termed ``massive cooling flow 
systems'').  CC systems with higher central cooling times 
(but less than about 4 Gyr) do exist (see, e.g., Fig.\ 7), but 
typically do not have published {\it Chandra} or {\it XMM-Newton} 
data available.  Therefore, as discussed in \S 4, use of 
equation (8) 
to represent observed CC clusters will yield the {\it minimum} 
amount non-gravitational heating needed to explain the 
structural properties of the CC cluster population.

Finally, in \S 5 we show that a plausible physical origin for 
this modification given by equation (8) could arise from the 
combined effects of preheating and radiative cooling.  It is 
important to note, however, that since the central cooling time 
of these CC systems is quite short this configuration is not 
stable.  Therefore, some form of episodic or continuous heating 
is required to maintain it.  As we demonstrate in \S 5, this can 
be achieved with AGN heating at the cluster centre.

\subsubsection{Scaled Entropy Profiles}

Gravitational shock heating naturally boosts the entropy of the 
intracluster gas at large radii to high values.  Under such 
conditions, any additional entropy/energy injection of the level 
normally associated with AGNs will only comprise a small 
perturbation.  In the limit where non-gravitational physics is 
negligible at large radii, gravitational self-similarity is 
expected to hold.  This implies that systems of varying total 
masses all have identical 
structure.  Therefore, once the total mass dependence of the 
various structural profiles (such as the temperature and 
density profiles) have been removed by dividing by some 
appropriate set of characteristic quantities (such as the virial 
temperature, some multiple of the critical density, etc.), the 
radial profiles of systems of varying mass will be equivalent.  
Comparing the profiles in these scaled units therefore tests 
further our understanding of the physics of the ICM.  We use the 
data of V06 and P06, which probe out to large radii, for this 
purpose.

For the observed systems, we scale the radial coordinate by the 
observationally-inferred estimates of $r_{500}$.  This quantity 
was estimated both by V06 and P06 (see also Pointecouteau et al.\ 
2005 in case of P06) by deriving the total mass profiles via a 
HSE analysis incorporating the spatially-resolved temperature and 
surface brightness profiles of the systems.  Both studies find 
that a NFW profile matches the inferred mass structure remarkably 
well and the resulting mass-concentration relationships are in 
excellent agreement with that derived from cosmological 
simulations and assumed by our models.  Self-similarity implies 
that the characteristic entropy of a system should scale with 
its temperature.  Thus, to scale the entropy of the observed 
profiles we make use of the measured mean 
spectral temperatures\footnote{We note that P06 have also 
presented their entropy profiles in scaled units (see their 
Figure 6) in order to make a direct comparison with the baseline 
entropy profile of Voit et al. (2005) (see equation 3).  However, 
this requires a significant extrapolation of the observed 
profiles, which typically can only reliably be measured out to 
$\approx r_{500} \approx 0.5-0.6 r_{200}$ (see Arnaud et al.\ 
2005).  Thus, there is some risk of introducing a bias by 
comparing the observed and theoretical properties out to $r_{200}$ 
(see, e.g., the discussion in the appendices of V06).  
Accordingly, we elect to compare the models to the data within 
$r_{500}$.}.  In both cases, the mean spectral temperatures have 
been ``cooling flow corrected'' by excising data from the central 
regions.  In the case of V06, the central (3-D) 70 kpc was 
excised, while in P06 (see also Arnaud et al.\ 2005) (projected) 
data within $0.1 r_{200}$ was excised.  In the models, we find 
that this mismatch results in only a few percent difference in 
the estimated mean spectral temperature.  We neglect this small 
difference.

\begin{figure}
\centering
\includegraphics[width=8.4cm]{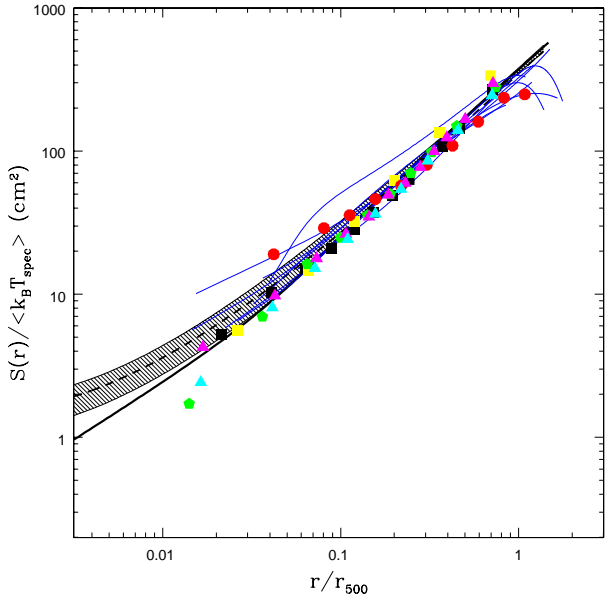}
\caption{This figure compares scaled entropy profiles, so that
the effect of cluster mass is removed. It is similar to Fig.~1
except that the entropy is normalised by the  ``cooling flow
corrected'' mean spectral temperature (see text) and the radial
coordinate is normalised by $r_{500}$. We omit the (small radii)
data of D06 since this scaling is only relevant outside the
cooling radius. Scaling the profiles in this way reduces the
scatter in the data at large radii.
 }
\end{figure}

To make a fair comparison between the models and the data, we 
compute the mean spectral temperature over a similar passband to 
that of {\it Chandra} and {\it XMM-Newton} (we use 0.1-10.0 keV) 
within the annulus $0.1 < r/r_{200} < 0.5$ for the model 
clusters.  This is done by fitting a single-temperature MEKAL 
plasma to the synthetic spectra extracted from this annulus.  We 
use the true (physical) value of $r_{500}$ to scale the radial 
coordinate for the model clusters.

We plot in Figure 2 a comparison of the scaled observed and 
theoretical entropy profiles.  Scaling according to the 
self-similar model has the effect of significantly reducing the 
scatter in the observed profiles.  In fact, the remaining scatter 
is consistent with statistical measurement uncertainties 
(typically of order 5-10\%), at least at large radii.  More 
importantly, the agreement between the observed and theoretical 
profiles, particularly at large radii, is exquisite.

Recently, there have been reports that the observed
entropy profiles at large radii are better described by a scaling
of $S \propto T^{2/3}$, as opposed to the self-similar
scaling $S \propto T$.  However, based on Fig.\ 2, such a scaling
does not seem to be required for high mass systems. (We remind the 
reader that we have omitted low mass systems with spectral 
temperatures less than 3 keV.)  The higher mass systems appear to 
obey self-similarity at large radii.

\subsubsection{Gas Density Profiles}

If the ICM is in HSE, then its density will be set by its entropy 
distribution and the depth of the (dark matter-dominated) 
gravitational potential well (see \S 2.1).  Therefore, a 
comparison between the observed and theoretical gas density 
distributions provides an excellent check of the results in \S 
3.1.1-3.1.2.

\begin{figure}
\centering
\includegraphics[width=8.4cm]{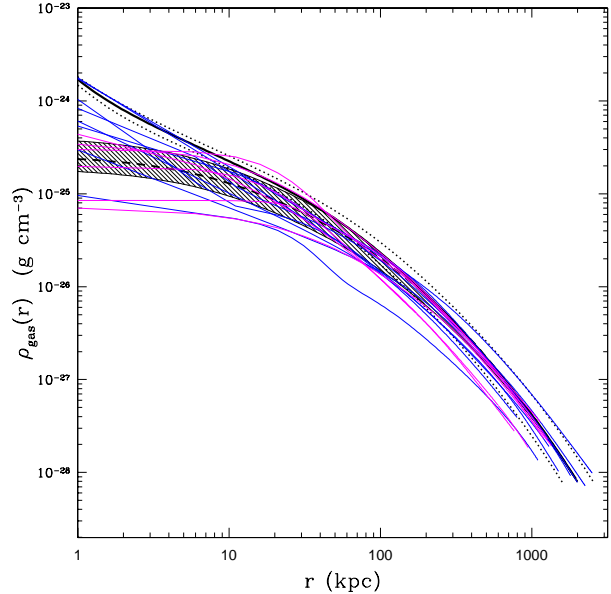}
\caption{Comparison of the {\it Chandra} gas density profiles of
V06 and the {\it XMM-Newton} gas density profiles of P06 with
the predictions of the pure cooling model.  The thin blue and
magenta curves represent fits to the density profiles of V06 and
P06, respectively.  The solid black curve represents the
predictions of the pure cooling model for a cluster with a
typical total mass $M_{200} = 10^{15} M_{\odot}$, while the two
dotted curves show the predictions for systems with $M_{200} =
5\times10^{14} M_\odot$ and  $2\times10^{15} M_\odot$ (bottom to
top).  The shaded region (scatter) and short-dashed line
(median) represent the linear modification to the pure cooling
entropy profile (see eqn. 8) for a cluster with $M_{200} =
10^{15} M_\odot$.
 }
\end{figure}

\begin{figure}
\centering
\includegraphics[width=8.4cm]{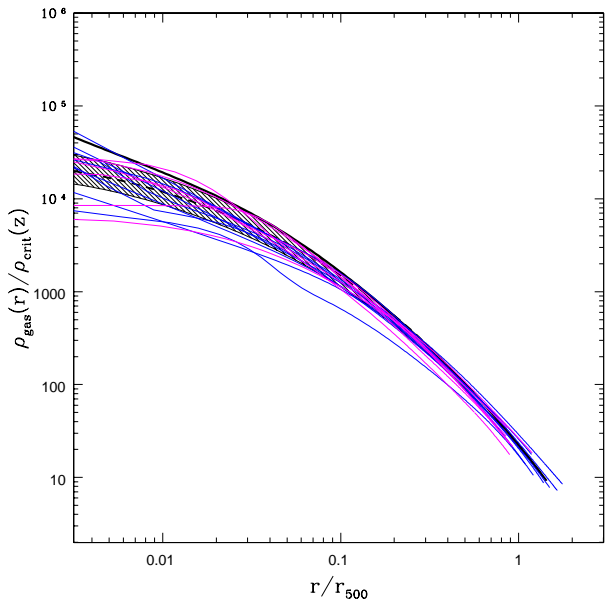}
\caption{Same as Fig.\ 3 except we scale the data and models
according to the gravitational self-similar model.  In
particular, the gas density is normalised by the critical
density of the universe at the redshift of the cluster and the 
radial coordinate by $r_{500}$.
 }
\end{figure}

In Figure 3, we plot the fits of V06 and P06 to their observed 
gas density profiles along with the gas density profiles of the 
model clusters plotted in Fig.\ 1.  A comparison of the observed 
gas density profiles with those predicted by the pure cooling 
model leads to the following conclusions: (1) at small radii the 
observed clusters are under-dense with respect to the pure 
cooling model; and (2) at large radii the observed and 
theoretical profiles show evidence for convergence.  If the pure 
cooling {\it entropy} profiles are modified  as in \S 3.1.1, we 
find that this yields gas density profiles in excellent
agreement with the observed profiles at small radii. 

Figure 4 presents the gas density profiles scaled according 
to the gravitational self-similar model.  In these scaled 
units the linear modification to the pure cooling model (see eqn. 
8) is in excellent agreement with the data over all radii.  
As noted above, this agreement is non-trivial.  Moreover, it 
confirms the results of \S 3.1.2, that the ICM at large radii 
scales according to the gravitational self-similar model.  If the 
entropy scaled as $S \propto T^{2/3}$, as recently proposed, this 
would imply $\rho_{\rm gas} \propto T^{1/2}$, but we see no 
evidence for this in the hot clusters plotted in Fig.\ 4.  

\subsubsection{Luminosity-Temperature Relation}

Up this point we have examined only the radial profiles of CC 
systems and their ability to be explained by a simple 
modification to the pure cooling model.  However, derivation of 
the deprojected (3-D) radial profiles is a model dependent 
process.  For example, spherical symmetry is often assumed as it 
considerably simplifies the analysis.  A single-phase medium is 
also implicitly assumed in this process.  The validity of these 
assumptions can be tested to some degree by comparing to the 
integrated luminosities and mean temperatures of clusters.  The 
idea is to compare the theoretical predictions to data that has 
undergone as little processing as possible.

Plotted in Fig.\ 5 is the bolometric X-ray luminosity---mean 
spectral temperature relation of clusters in the {\it ASCA} 
Cluster Catalog of Horner (2001).  Importantly, neither the 
luminosities nor temperatures of the observed systems have been 
corrected for the presence of cool cores.  As noted in previous 
studies (e.g., Markevitch 1998; M04, O'Hara et al.\ 2006; Balogh 
et al.\ 2006, Chen et al.\ 2007), the scatter in the 
(uncorrected) observed 
luminosity-temperature relation cannot be accounted for by 
measurement uncertainties.  Rather it is due mainly to 
variations in the properties of the ICM in the cores of 
clusters.  This is clearly demonstrated by colour coding a 
subset of the clusters according to their 
central cooling time, which we estimate using the central {\it 
ROSAT} gas density measurements Mohr et al.\ (1999) (kindly 
provided by J. Mohr) and the ACC temperatures.  Clearly 
there is a strong trend in Fig.\ 5, with clusters that have 
short (long) cooling times being over-(under-)luminous with 
respect to the mean relation.

Consistent with the results presented above, we find that the 
observed systems are under-luminous (because they are 
under-dense) with respect to the pure cooling model (see 
also Voit et al.\ 2002; M04).  Modifying the entropy profile at 
small radii as done in \S 3.1.1 brings the theoretical 
predictions into excellent agreement with the observed relation 
of CC systems.  We discuss the NCC systems plotted in Fig.\ 5 in 
\S 3.2.

\begin{figure}
\centering
\includegraphics[width=8.4cm]{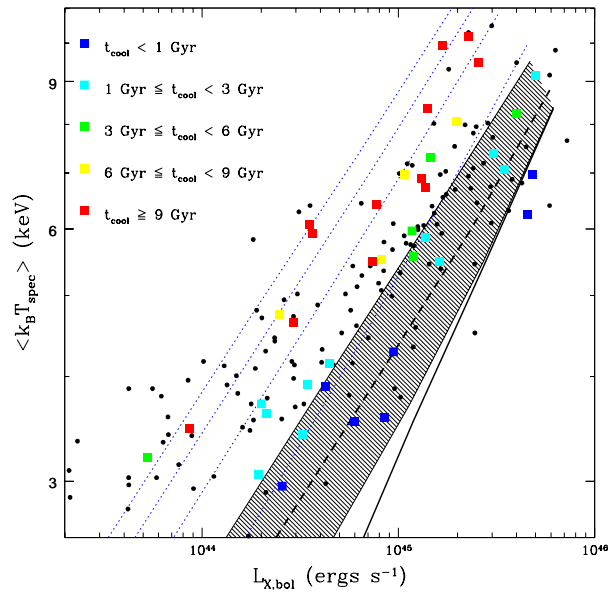}
\caption{Comparison of the {\it ASCA} bolometric luminosity-mean
spectral temperature relation of Horner (2001) with the
predictions of the pure cooling model.  The data points represent
the $L-T$ measurements of Horner (2001).  The solid black
curve represents the predictions of the pure cooling model.
The shaded region (scatter) and short-dashed line
(median) represent the linear modification to the pure cooling
entropy model (see eqn. 8).  The thin blue dotted lines
represent pedestal modifications to the pure cooling model with 
amplitudes of $100$, $300$, $500$, and $700$ keV cm$^2$ (bottom 
to top; see \S 3.2).
 }
\end{figure}

\subsubsection{Spectral Properties}

A further important observation that must be reproduced is
the absence of many low-energy iron lines in the {\it XMM-Newton} 
RGS spectra of CC systems.  As discussed in Peterson et al.\ 
(2003) (hereafter P03), a number of transitions in Fe provide a 
very sensitive thermometer of low temperature ($0.2$ keV $< k_B T 
< 4$ keV or so) plasmas. P03 compared their RGS spectra of a 
sample of CC systems with the predictions of the cooling flow 
model of Fabian and collaborators (e.g., Fabian \& Nulsen 
1977; Fabian et al.\ 1984; Johnstone et al.\ 1992; 
see also Cowie \& Binney 1977) and concluded that there is a 
significant deficit of emission from many key low-energy iron 
lines (see also Peterson et al.\ 2001; Kaastra et al.\ 2004).  
The implication is that the cooling of the ICM appears to be 
halting at a temperature floor of a few keV.  

\begin{figure}
\centering
\includegraphics[width=8.4cm]{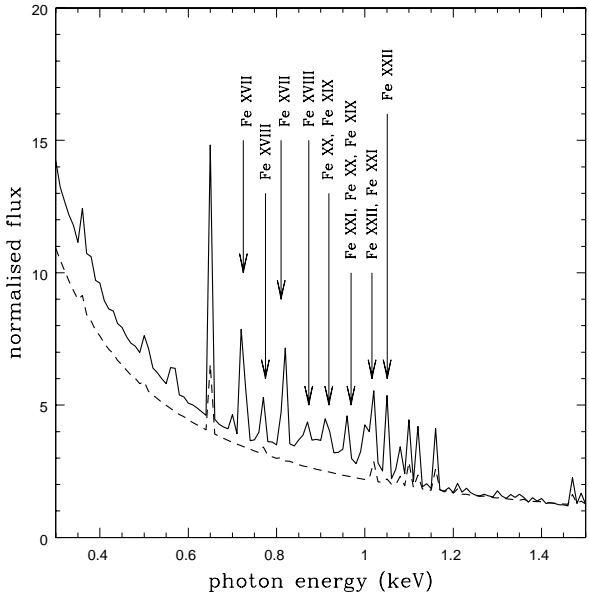}
\caption{The predicted X-ray spectra of the pure cooling model
(solid curve) and the linear modification to pure cooling 
model (short-dashed curve).  The spectra have been extracted from
the central (projected) 100 kpc of a $M_{200} = 10^{15} M_\odot$
model cluster and have been normalised to match at high photon
energies (not shown).
 }
\end{figure}

Given the excellent simultaneous match of the modified pure 
cooling model to the entropy and gas density profiles, one 
might expect automatic agreement with the X-ray spectra.  
However, it should be kept in mind that: (1) the P03 cluster 
sample differs from the samples of V06 and P06; (2) the 
instrument characteristics (and systematics) of the {\it 
XMM-Newton} RGS are quite different from those of the {\it 
Chandra} ACIS or {\it XMM-Newton} EPIC instruments; and (3) in 
the case of V06 and P06 single-phase spectral models were 
assumed and the profiles were azimuthally-averaged.  This last 
point is worth particular attention.  Unlike the other 
instruments, which have comparatively poor spectral resolution, 
the {\it XMM-Newton} RGS can resolve the low-energy Fe L 
complex.  Consequently, if there is a significant amount of cool 
gas (below a few keV) it will be detected, regardless 
of whether the gas is single- or multi-phase or if there has 
been azimuthal averaging.  Thus, an examination of the agreement 
(or lack of) between the RGS spectra and the models is 
worthwhile.

Plotted in Figure 6 is the predicted low-energy X-ray spectra of 
the pure cooling model and the linear modification to the 
pure cooling model.  The various spectral lines all probe 
slightly different temperature ranges, but the ones to focus on 
are Fe XXII, Fe XXI, Fe XX, Fe XIX, Fe XVIII, and Fe XVII (which 
are labelled), all of which are sensitive to  temperatures of 
$k_BT < 1.5$ keV (see Table 1 of P03).  These lines are 
completely absent in the RGS spectra of the moderate mass ($4$ 
keV $< k_BT_{\rm spec} < 7$ keV) CC clusters studied by P03, and 
show up only weakly in their lower mass systems.

A comparison of the predicted spectra clearly demonstrates 
that the relevant lines are virtually absent when the pure 
cooling model is modified by eqn. (8).  The simple reason for 
this behaviour is that boosting the central entropy has the 
effect of boosting the 
central temperature of the ICM.  For our model clusters spanning 
the mass range $5\times10^{14} M_\odot \leq M_{200} \leq 
2\times10^{15} M_\odot$ we find minimum (3-D) temperatures 
ranging from $2.1$ keV $\leq k_BT \leq 3.4$ keV, with the highest 
minimum temperatures corresponding to the most massive systems.  
Phrased in another way, we find that the coolest gas in the 
modified pure cooling model is only $2-3$ times lower than the 
ambient ICM temperature.  This agrees very well with the 
observational estimates of P03.

The simultaneous agreement with the RGS spectra and the gas 
and density profiles therefore points to a simple picture for 
the ICM in CC systems; it is fully consistent with a single-phase 
medium in hydrostatic equilibrium within a gravitational 
potential whose shape was predicted by cosmological simulations.  
(In Appendix B, we go further and show that previous arguments 
used in support of multi-phase cooling are, in fact, fully 
consistent with our single-phase models.)  In \S 5, we describe a 
simple physical model that can give rise to the appropriate 
entropy distribution of CC clusters.

\subsection{Non-cool core (NCC) clusters}

There is currently a dearth of high spatial and spectral 
resolution observations of NCC clusters, even though they could 
represent more than half of all clusters.  We therefore compare 
our models to available {\it ROSAT} and {\it ASCA} data.

In terms of radial profiles, only the gas density profile (as 
opposed to the temperature or entropy profile) can be robustly 
determined in the absence of high quality spatially-resolved 
spectra.  The gas density profile can essentially be derived 
from the X-ray imaging alone, since the X-ray luminosity scales 
as $\rho_{\rm gas}^2 \Lambda(T) \approx \rho_{\rm gas}^2 
T^{1/2}$.  So even if one assumes an ICM temperature that is 
incorrect by a factor of 2 this translates to less than a 20\% 
error in the gas density estimate.  This is good news since 
theoretical models predict (and new spatially-resolved 
observations confirm, at least in the case of CC systems) that 
the temperature of the ICM varies by less than about a factor of 
2 from the cluster centre to its periphery.

Plotted in Figure 7 are the {\it ROSAT} gas density profiles of a 
sample of 45 hot clusters (selected from the 55 clusters in the 
flux-limited sample of Edge et al.\ 1990) studied by Mohr et 
al.\ (1999; kindly provided by J. Mohr).  These profiles were 
derived assuming an isothermal ICM using mean spectral 
temperature measurements from the literature.  We have adjusted 
the gas density profiles to conform with the more recent 
(uniform) {\it ASCA} temperature measurements of Horner (2001) 
but, for the reason mentioned above, this is only a very minor 
correction.  The radial coordinate has been normalised by 
$r_{500}$, where this quantity has been estimated using the 
observationally-calibrated relation (see Mohr et al.\ 
1999):

\begin{figure}
\centering
\includegraphics[width=8.4cm]{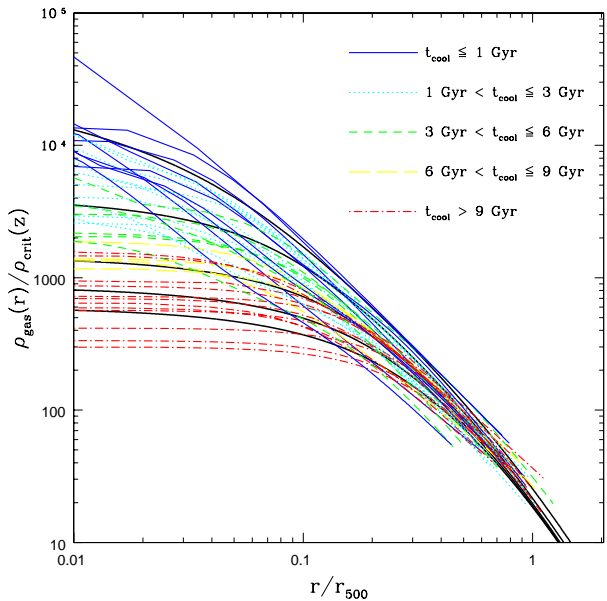}
\caption{
Comparison of the {\it ROSAT} gas density profiles of Mohr et
al.\ (1999) with the predictions of the pure cooling model.  The
dashed lines represent the data of Mohr et al.\ (kindly provided
by J. Mohr).  They have been colour-coded according to estimated
central cooling time using the same colour scheme as in Fig.\ 5.
Under this scheme,  the green, yellow, and red curves correspond
to NCC clusters. The radial
coordinate has been normalised by $r_{500}$, where this quantity
has been estimated for both the data and models using an
observationally-calibrated relation between $r_{500}$
and $<T_{\rm spec}>$ (see eqn.~9).  The solid black curves 
show the effect of modifying the pure cooling model with pedestals 
of amplitude 10, 100, 300, 500, and 700 keV cm$^2$ (top to 
bottom). The models shown are for the case $M_{200} = 10^{15} 
M_{\odot}$.
 }
\end{figure}

\begin{equation}
r_{500} = 1.19 h^{-1} \ {\rm Mpc} \ \biggl[\frac{<k_BT_{\rm 
spec}>}{10 \ {\rm keV}} \biggr]^{1/2}
\end{equation}

At small radii, the observed gas density profiles show more than 
an order of magnitude in dispersion, with the scatter clearly 
being strongly correlated with the estimated central cooling 
time and therefore CC/NCC designation.  The bulk of the 
dispersion at small radii, however, is associated with the NCC 
population.  At large radii (beyond $\approx 0.2 r_{500}$), the 
gas density profiles tend to converge and there is no strong 
evidence for a correlation between central cooling time and the 
dispersion in the gas density there.  In other words, outside the 
core the ICM properties of NCC and CC clusters are very similar, 
as noted previously (e.g., Markevitch 1998; Neumann \& Arnaud 
1999).  

In the absence of high quality data to actually measure 
$S(r)$ for the NCC systems in which we are interested here, we 
assume that their entropy structure can be explained by adding 
a pedestal (i.e., a constant value to all radii) to the pure 
cooling entropy profile, and we constrain the range of amplitudes 
necessary to explain the observed gas density profiles.  Note 
that this procedure implicitly assumes that the ICM in NCC 
systems is similar to that in CC systems; i.e., it is a 
single-phase medium in HSE confined within a similar NFW  
gravitational potential.  Detailed measurements are obviously 
needed to test these assumptions.

In agreement with our previous findings, adding a pedestal of 
$\approx 1-10$ keV cm$^2$ to the entropy profile of the pure 
cooling model results in a scaled gas density profile that is in 
reasonably good agreement with the extreme CC systems in 
Fig.\ 7 
(see \S 3.1.1).  Although from the comparison with higher 
quality data we can expect the linear modification (see eqn. 8) 
would yield even better agreement with these systems, since our 
focus here is on the NCC systems where such high resolution data 
is unavailable, use of this more complex modification is 
unwarranted.

\begin{figure}
\centering
\includegraphics[width=8.4cm]{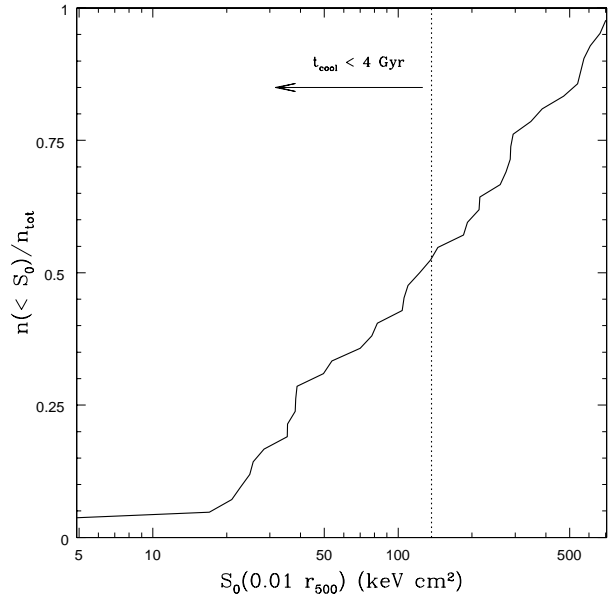}
\caption{The expected distribution for the central entropy of 
clusters, derived by placing the observed {\it ROSAT} gas density 
profiles of Mohr et al.\ (1999) (for clusters from the 
flux-limited sample of Edge et al.\ 1990) in hydrostatic 
equilibrium within NFW gravitational potentials.  If these 
assumptions are correct, this is the distribution we expect to be 
shown by upcoming representative {\it Chandra} and {\it 
XMM-Newton} samples.  The dotted vertical line delineates 
systems with implied central cooling times of greater or less 
than 4 gigayears.  We note that the central entropy 
distribution of flux-limited (as opposed to mass-limited) samples 
could be biased somewhat low.}
\end{figure}

Modifying the pure cooling entropy profile with pedestals of 
amplitudes ranging from approximately $100$ keV cm$^2$ to 700 
keV cm$^2$ brackets the gas density profiles of most of the 
observed NCC systems.  Not only does the pedestal modification 
with this range of amplitudes simultaneously match the intrinsic 
scatter at small and large radii, but it also does well in 
matching the shapes of individual profiles.  

A look back at Fig.\ 5 reveals that this range of amplitudes 
also brackets the NCC systems on the observed 
luminosity-temperature relation.  (The predicted slope of 
the $L-T$ relation for a model with fixed pedestal amplitude is 
steeper than the observed relation, but this is not of concern 
since the observed scatter and Fig.\ 7 clearly rule out a 
universal pedestal amplitude.) In fact, the inferred range of 
pedestal amplitudes necessary to explain the data plotted in 
Figs.\ 5 \& 7 is in excellent agreement with constraints 
based on studies of the scatter in various other X-ray and SZ 
effect scaling relations (e.g., Babul et al.\ 2002; McCarthy et 
al.\ 2002; 2003; M04; Balogh et al.\ 2006).

Finally, in anticipation of the release of {\it 
representative} samples of clusters observed at high resolution 
(e.g., with {\it Chandra}), we present in Fig.\ 8 our 
expectation for the distribution of the central entropies in
these clusters.  The central entropies are derived by 
placing the {\it ROSAT} gas density profiles of Mohr et al.\ 
(1999) in hydrostatic equilibrium within a NFW gravitational 
potential.  The central entropies, $S_0$, are measured at a 
fiducial radius of $0.01 r_{500}$, typically corresponding to 
a physical radius of $\approx 10-15$ kpc.  Our results are not 
sensitive to this choice of radius, since almost all the 
clusters show evidence for cores at least this size. 

Fig.\ 8 shows that only 5\% of systems should have central 
entropies of $\approx 10$ keV cm$^2$ or lower.  This puts into 
perspective the focus of the majority of observational studies 
based on {\it Chandra} and/or {\it XMM-Newton} data so far.  
Furthermore, we find that 25\%, 50\%, and 75\% of clusters should 
have central entropies that are approximately less than 40, 120, 
and 290 keV cm$^2$, respectively.  Note that, at least in 
the models, systems with central entropies of $S_0 < \ \approx 
130$ keV cm$^2$ have $t_{\rm cool} < \ \approx 4$ gigayears and 
would therefore be classified as CC clusters.

To a good approximation, we find that the distribution plotted 
in Fig.\ 8 can be represented by the following functional form:

\begin{equation}
\frac{n(< S_0)}{n_{tot}} = 0.57 \log_{10}{\biggl(\frac{S_0}{1 
\ {\rm keV \ cm}^2} \biggr)} - 0.67
\end{equation}

\noindent for $20$ keV cm$^2$ $< S_0 < 850$ keV cm$^2$.

Note that, given the current data, this is the distribution that 
should result if NCC clusters are dominated by NFW profiles, with 
single-phase gas in hydrostatic equilibrium, as the CC clusters 
appear to be.  If the upcoming studies reveal a distribution that 
is significantly different from this one, it might signal a break 
down of one or more of these assumptions for the NCC population.

\section{Heating Energy Requirements}

To explain the observed properties of the ICM, energy is required 
to (1) transform the initial pure cooling state to the observed 
one; and to (2) maintain this state. We demonstrate below that 
in order to satisfy the first condition, huge amounts of energy 
are required if the systems are heated from within at (or near) 
the present day.  This is especially the case for the NCC 
systems.  Preheating is an attractive, energy-efficient 
alternative for satisfying the first condition.  We then 
demonstrate that once the observed state has been achieved, 
maintaining this configuration is energetically much easier to 
accomplish.  This energy might, for example, come from jets or 
bubbles injected by AGN at the cluster centre, although the
details of the energy injection are not pertinent to the 
argument. 

\subsection{Reaching the observed state with internal heating}

If the ICM evolves initially to the convergent pure cooling state 
and then is heated from within at (or near) the present 
day, the total energy that must be injected in order to
explain the structure of clusters is just the difference of
the total energies of the observed systems and the pure cooling 
model.  We use the modifications to the pure cooling 
model discussed in \S 3.1-3.2 to represent ``smoothed'' versions 
of the observational data.  Note that this energy calculation 
does not make any assumptions about the form of the heating.  In 
the absence of sources or sinks, there is a unique 
(path-independent) energy requirement to transform the 
properties of a gas from some initial configuration to some 
final configuration.  In the case of the ICM, radiative 
losses that take place after (or during) the heating are likely 
relevant, which implies that the above calculation actually 
yields a {\it minimum} energy requirement.  However, the 
structural uniformity of the gas at small radii in extreme 
CC clusters (e.g., all have minimum entropies of $\lesssim 10$ 
keV cm$^2$) combined with fact that the majority of time spent 
cooling is at the highest entropies (lowest densities) argues 
against energy injection levels significantly higher than 
inferred with the above approach, at least for the CC systems in 
\S 3.1.  In the case of extreme NCC systems, radiative losses 
should be unimportant and therefore our energy estimates should 
be robust for these systems as well.  It is the systems that lie 
in between these two extremes where our energy estimates could 
be underestimated by up to a factor of two or so.  

\begin{figure}
\centering
\includegraphics[width=8.4cm]{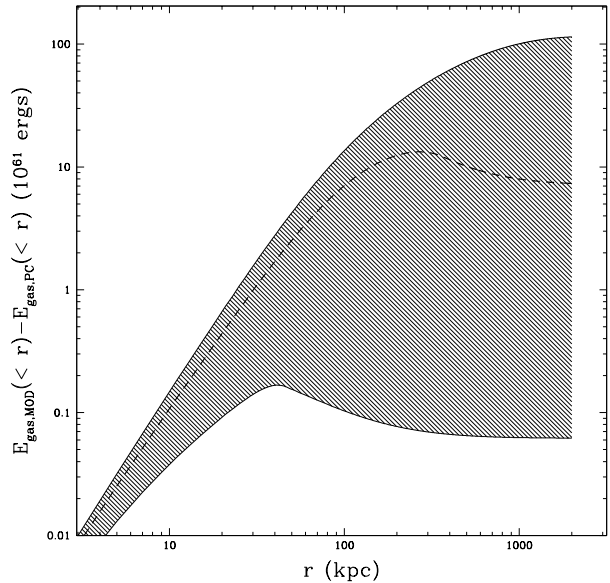}
\caption{Cumulative energy distribution requirements for
transforming from the pure cooling state to the observed CC state
for a typical $M_{200}=10^{15} M_\odot$ cluster.  The shaded 
region (scatter) and short-dashed curve (median) represent the 
linear modification to the pure cooling model (see eqn. 
8).  We note that raising the entropy of the central gas, 
while decreasing the gas density at small radii, can sometimes 
slightly increase the density at intermediate radii.  This, in 
turn, can slightly increase the potential energy of the gas, 
giving rise to a non-monotonic energy deposition vs. 
radius curve.}
\end{figure}

In Figure 9, we present the cumulative energy distributions 
required to transform from the pure cooling state 
to the observed state of CC clusters. Evidently, a total 
energy ranging from $\sim 10^{60}-10^{63}$ ergs, with a typical 
value of $\sim 10^{62}$ ergs, must be injected into the ICM to 
explain the structural properties of cool core clusters.  It is 
interesting to note that an energy of $\sim 10^{62}$ ergs 
is roughly equivalent to 10\% of the total thermal energy of a CC 
cluster (within $r_{200}$) or half the total energy a CC cluster 
would radiate over a period of 10 gigayears.  But such a large 
injection of energy is not beyond the realm plausibility.  For 
example, the recently discovered ``cluster-scale'' X-ray bubbles 
in Hercules A (Nulsen et al.\ 2005) and MS0735.6+7421 (McNamara 
et al.\ 2005), which were presumably blown by a central AGN, 
require up to $6\times10^{61}$ ergs to inflate.  This is only a 
factor of a few lower than required to account for CC systems.  
However, such systems appear to be rare at the present epoch.

In order to explain the observed structure of NCC systems, we 
find that typically an order of magnitude more energy is required.  
In particular, total energies ranging between $\approx 1-4 \times 
10^{63}$ ergs are needed, where the lower and upper values 
correspond to the 100 keV cm$^2$ and 700 keV cm$^2$ pedestal 
modifications, respectively.  This dwarfs the energy output for 
the most powerful AGN bursts known in Hercules A and 
MS0735.6+7421.  But McCarthy et al.\ (2007b) have shown that such 
large energy requirements are still within the range of 
theoretical plausibility 
for AGN if one adopts an efficiency factor of $\sim0.1$, as 
predicted by jet models with near-maximally rotating black holes 
(e.g., Nemmen et al.\ 2007), and takes into account outflows 
from not only the central BCG but from all the bulge-dominated 
systems in the cluster (e.g., Nusser et al.\ 2006).  On the 
other hand, the above calculations would imply that {\it every} 
cluster has had outbursts at least as powerful as those observed 
in Hercules A and MS0735.6+7421 at some point during their 
lifetimes and perhaps 50\% of those have had bursts that were an 
order of magnitude larger still.

\subsection{Reaching the observed state with preheating}

In \S 4.1, we assumed that the ICM reached the pure cooling
configuration and was then heated by some means to the observed 
configuration.  The energy requirements to reach the observed 
configuration can potentially be lowered, however, if the 
intracluster gas is heated before it falls into the cluster 
potential well (e.g., Kaiser 1991; Evrard \& Henry 1991; 
Bower 1997; Balogh et al.\ 1999; Babul et al.\ 2002; Voit et al.\ 
2002; Oh \&  Benson 2003).  Such ``preheating'' is much more 
energetically efficient than internal heating (see Appendix 
C for a detailed comparison between the energetics of preheating 
and internal heating).  Using simple 1-D models, 
Voit et al.\ (2003) have shown that if the accretion is smooth, 
preheating adds a pedestal to the baseline gravitational entropy 
profile with an amplitude that is basically the preheating 
level, $S_{\rm PH}$ (see also Lu \& Mo 2007).  Assuming the 
gas is initially cold, and therefore its thermal energy prior to 
preheating is negligible, the energy required to add a pedestal 
of amplitude $S_{\rm PH}$ to the baseline gravitational entropy 
profile\footnote{Note this is essentially equivalent to adding a 
pedestal to the pure cooling entropy profile so long as the 
level of preheating is large compared to the difference between 
the baseline gravitational and pure cooling entropy profiles.  
This condition is easily satisfied for physically interesting 
values of $S_{\rm PH}$.} is given by:

\begin{eqnarray}
E_{\rm PH} & = & \frac{3}{2} \ f_b \ M_{200} \ m_p^{5/3} \ S_{\rm 
PH} \\
 & & \ \times \ \ [\mu \mu_e \Omega_b \rho_{\rm 
crit}(z=0)(1+z)^3]^{2/3} \nonumber \\
 &  \approx & 2.0 \times 10^{60} \ {\rm ergs} 
\ \biggl(\frac{f_b}{0.13} \biggr) \biggl(\frac{M_{200}}{10^{15} 
\ M_\odot} \biggr) \nonumber\\
 &  & \ \ \times \ \ \biggl(\frac{S_{\rm PH}}{100 \ {\rm keV \ 
cm}^2} \biggr) \biggl(\frac{\Omega_b h^2}{0.02} \biggr)^{2/3} 
(1+z)^2 \nonumber
\end{eqnarray}

The energy required to preheat the ICM therefore scales linearly 
with the preheating entropy and with redshift as $(1+z)^2$.  The 
redshift dependence enters through decreasing baryon density of 
the universe due to Hubble expansion.  We note, however, that 
equation (11) will underestimate the required energy somewhat 
at low redshifts.  This is because the density of the baryons 
will likely exceed the universal mean value $\rho_b(z) = 
\Omega_b \rho_{\rm crit}(z)$, since some of the baryons will 
have begun to collapse and will have decoupled from the global 
expansion of the universe (see, e.g., Balogh et al.\ 1999).  In 
addition, some fraction of these baryons may already have 
collapsed into smaller dense (possibly virialised) subsystems 
(e.g., Voit et al.\ 2003). Equation (11) therefore represents 
the minimum energy required to preheat the gas to entropy 
$S_{\rm PH}$ at redshift $z$.  

\begin{figure}
\centering
\includegraphics[width=8.4cm]{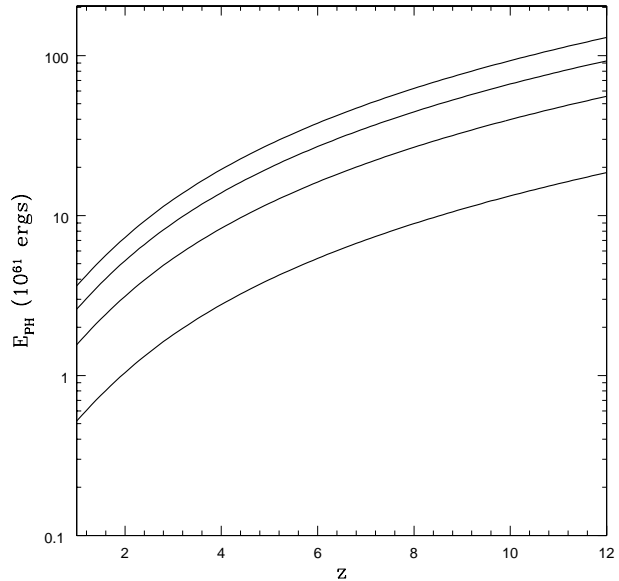}
\caption{
Preheating energy requirements for a typical $M_{200} = 10^{15}
M_\odot$ cluster.  Shown bottom to top are the predictions of
eqn. (11) for preheating levels of $S_{\rm PH} =$ 100, 300,
500, and 700 keV cm$^2$, respectively.
 }
\end{figure}

The relevant range of redshifts would presumably fall between 
the epochs of cluster formation and reionization, which is when 
the first sources of non-gravitational heating turned on.  A 
physically sensible range would therefore be $1 < z < 12$.  
In Fig.\ 10 we plot the energy required to preheat the ICM of a 
typical $M_{200} = 10^{15} M_\odot$ cluster as a function of 
redshift for various choices of $S_{\rm PH}$.  As we discuss 
later (see \S 5), for clusters preheated to less than about 
$300$ keV cm$^2$, there is sufficient time between the epoch of 
preheating and the present day for radiative cooling losses to 
reduce the amplitude of the entropy core to the level observed 
in present-day CC clusters.  Clusters preheated above this level, 
however, will not have had the time to remove their 
initial entropy cores.  Such systems could potentially provide a 
viable explanation for present-day NCC  clusters.  

We find that if preheating occurs late, say from $1 < 
z < 2$ (i.e., approximately when the universal star formation 
rate peaks), one can plausibly explain the present-day structure 
of both CC and NCC clusters with up to two orders of 
magnitude less energy than if these systems were heated from 
within at the present-day.  For example, a Hercules A-like 
outburst at $z \sim 1-2$ could potentially 
account for even the most extreme NCC clusters today.  However, 
there is little to be gained if preheating occurs close to the 
epoch of reionization, simply because the baryon density is much 
higher then and therefore much greater energy is required to 
raise the entropy of the gas to the required level.

While (late) preheating is a physically attractive scenario, 
whether or not it actually occurred is still not known. At
present, there is no theory that self-consistently links
preheating to galaxy and/or black hole formation. There 
are several pieces of circumstantial evidence in support of 
preheating.  For one, there are no {\it observed} sources of 
non-gravitational heating powerful enough to explain the 
structure of moderate and extreme NCC clusters if heated from 
within.  Secondly, it is now well known that AGN, the most likely 
culprit behind preheating, were much more active in the past than 
they are at present-day (e.g., Schmidt \& Green 1983; Hasinger et 
al.\ 2005).  Finally, the observed diffuse X-ray background 
appears to be too faint compared to that predicted by 
cosmological simulations that do not include some very energetic 
and widespread form of non-gravitational heating (Pen 1999; 
Bryan \& Voit 2001; but see Roncarelli et al.\ 2006).

On the other hand, high resolution {\it Chandra} data (e.g., D06; 
Sanderson et al.\ 2006) indicates that the central cooling time 
of the gas in many CC clusters typically ranges from $t_{cool} 
\approx 1-4 \times 10^8$ yr.  The abundance of such systems 
makes it highly unlikely that preheating is the whole story for 
these CC systems.  It would be a remarkable coincidence if such 
clusters were (pre)heated only once and that now we just happen 
to be observing them a mere $10^8$ yrs before the onset of 
catastrophic cooling.  The same argument also applies in the 
case of heating following cluster formation --- a single burst 
seems irreconcilable with the short central cooling time of the 
ICM in CC clusters.  It therefore seems inevitable that
there is either episodic or continuous heating occurring in CC 
clusters at the present day.   But, as we demonstrate in \S 5, 
this does not imply that preheating does not have a role to play 
in the formation of CC systems.  

In the case of NCC clusters, the central cooling times are many 
gigayears (by definition), so a single bout of preheating 
is fully compatible with the current suite of observational 
data (see M04 for additional discussion).

\subsection{Maintaining the observed state}

Maintaining the current (observed) structural properties is 
energetically much easier than getting to this state in the 
first place.  For example, within the central 300 kpc, where the 
pure cooling model deviates from the observed systems, a typical 
$M_{200}=10^{15} M_\odot$ CC cluster radiates $\approx 2 \times 
10^{45}$ ergs s$^{-1}$.  The central cooling time of the gas 
in CC clusters is $t_{\rm cool} \sim 10^8$ yr, implying that  
non-gravitational heating events must be occurring on this 
timescale or shorter.  There is mounting observational evidence 
that suggests AGN-blown bubbles are somehow offsetting radiative 
losses in CC clusters.  Typical (i.e., not cluster-scale) X-ray 
bubbles in massive clusters require $\sim 10^{60}$ ergs to 
inflate (e.g., B{\^i}rzan et al.\ 2004) which, in order to 
balance radiative losses, must be dissipated on a timescale of 
$\sim 10^{7}$ yr.  This is consistent with the cooling time 
estimates and also with the estimated ages of observed bubbles 
(e.g., Dunn \& Fabian 2006), which is suggestive that the 
bubbles could indeed be responsible for offsetting radiative 
cooling losses.

It would be possible to stop the paper at this point, and 
conclude that, while preheating must play a role in NCC 
clusters, the energetics of bubble heating appear sufficient to 
offset the cooling luminosity of CC clusters.  However, it has 
proved difficult to construct a simple, steady-state model for 
CC clusters in this way, since it is then necessary to 
efficiently distribute this energy over the entire cooling region.
In the following section we demonstrate that preheating  
alleviates this necessity, by greatly delaying the cooling
of CC clusters, holding them in their observed states for 
longer than the lifetime of the systems.   Moreover, this 
allows us to construct a single framework for explaining both 
NCC and CC clusters.  We emphasise, however, that the 
conclusions we have reached in this section do not depend at 
all on the particular heating model we describe below.

\section{A SIMPLE HEATING MODEL FOR COOL CORE CLUSTERS}

Clearly preheating cannot be the whole story 
for CC clusters.  Given the relatively short central cooling 
times in clusters preheated to $S_0 \lesssim 300$ keV cm$^2$, 
some form of ongoing heating is also needed.  Bubbles inflated
by AGN have the energy required to do this; the difficulty lies
in coupling this energy to the cooling gas. Conventionally, the
heat input is pictured as balancing the cooling emissivity 
throughout the cooling region. If heating matches cooling
shell by shell, there is no flow of gas and the cluster will
exist in a steady state. The problem with this approach is that
it is difficult to imagine mechanisms where the radial distribution
of the heat input is naturally tuned to match the radial
dependence of the cooling rate. Heating is typically a 
$\rho$-dependent process, while collisional cooling depends on 
$\rho^2$.  For example, in Appendix D, we outline the 
difficulties of distributing heat through thermal conduction: 
without ad-hoc tuning of the magnetic field structure, heating 
the centre of the cluster actually leads to increased cooling 
at intermediate/large radii.

We therefore take another approach to this problem: an 
acceptable solution does not necessarily require a balance 
between heating and cooling within each shell that maintains a 
perfect steady state.  Indeed, it is acceptable to 
preferentially heat the cluster at its centre so 
long as the resulting gas flows preserve the system's density 
profile, only allowing the profile to evolve over timescales 
longer than the lifetime of the system (i.e., on timescales $>5$ 
gigayears). Below we consider an extreme example of such a 
scenario. By heating the gas only at the cluster centre and 
then allowing it to float buoyantly outwards, we show that a 
quasi-steady state can be established. By combining this heating 
model with the preheating that we have discussed in \S4, we show 
that this model provides a good explanation of CC cluster 
properties.

\subsection{The ``wood stove'' approximation}

A wood stove is able to heat your house even though it is small
and located in the corner of a room. It does this by heating the 
air in its immediate surroundings. The warm air convects around 
the house mixing with the cooler air and distributing its heat.
It works because of the buoyancy of the hot air compared to the 
ambient temperature.  Compared to an open fire, relatively 
little heat is radiated, making the stove vastly more efficient. 
This type of heating is somewhat analogous to the picture of 
AGN heating that we are presenting.  In our model, we are 
assuming that the radiant heating (referred to as ``distributed 
heating'' in clusters) is negligible.  Because it heats at a 
lower temperature, an immersion water heater is possibly a 
better, but less poetic, analogy.

\subsubsection{Model description}

In the following sections, we consider the effect of embedding 
an AGN ``wood stove'' within a cluster. We first consider a 
cooling cluster that has not been preheated. This is essentially 
a simplified version of the detailed ``circulation flow'' models 
explored by Mathews et al.\ (2003; 2004).  The treatment 
described below is incorporated into our time-dependent hydro 
algorithm outlined in \S2.  First, we specify a minimum entropy 
threshold that signals when the central AGN is to `switch on'.  
So when radiative cooling reduces the entropy of a parcel of gas 
below this threshold, we assume that it is heated by the AGN.  We 
use the observations of D06 as a guide and adopt a minimum 
entropy threshold of 10 keV cm$^2$ (note, however, that the 
minimum entropy need not be fixed to match these specific data).
As in Mathews et al. (2003; 2004), we do not specify the physical 
mechanism that is responsible for heating the gas at small 
radii.  One possibility is that outflowing bipolar jets 
shock heat the gas at small radii, so long as they do not 
introduce large entropy inversions in the ambient ICM (Voit 
\& Donahue 2005).  Another possibility is heating by cosmic 
rays associated with the jet.  Since the entropy profiles of the 
model clusters increase monotonically with radius, the heating 
always occurs near the cluster centre, at radii of a few kpc.  
The minimum entropy threshold of 10 keV cm$^2$ 
corresponds to the minimum entropy that the gas parcel can be 
heated to, $S_{\rm H,min}$.  The maximum entropy that the parcel 
of gas can be heated to, $S_{\rm H,max}$, is left as a free 
parameter which we vary.  As described below, this is equivalent 
to varying the maximum (3-D) radius a bubble can float to.  

Physically, there is no reason to expect that successive AGN 
bursts will always heat gas to the same entropy.  Variations are 
to be expected as the local properties of the ICM are time 
variable.  However, it is beyond the scope of the present study 
to self-consistently link the properties of the cooling ICM with 
the accretion disk physics of black holes.  Instead, we simply 
assume a uniform probability heating distribution between $S_{\rm 
H,min}$ and $S_{\rm H,max}$, so that any value within this range 
is equally likely to result.  Once a new entropy has been 
randomly assigned, we assume the parcel of gas floats 
adiabatically out to the radius where its entropy matches that of 
the ambient ICM.  This is the termination radius, where the 
material is deposited.  

In order to clearly distinguish our model from conventional
``distributed'' heating scenarios, we assume that the bubbles
do no heating as they rise (see Appendix E for further 
discussion). In 
order to further simplify the calculations, we assume that the 
bubbles convect out to their isentropic surface instantaneously.  
One could instead properly take into account the rise time of 
the material by computing the drag force exerted on the rising 
material by the ICM if the bubble size is known.  However, 
since the entropy of the material is typically much higher than 
that of the surrounding ICM (except when it approaches the 
termination radius), this implies that its density is much 
lower\footnote{We assume the bubbles remain in pressure 
equilibrium with the ambient ICM as they rise.} (this is why 
the bubbles appear as depressions in X-ray images). 
Consequently, the material will not contribute much to the 
observed X-ray properties of clusters, so long as the volume 
occupied by the bubbles is relatively small.  For this reason, 
one can justifiably neglect the contribution of the hot rising 
material to those properties of interest here, such as the 
predicted entropy profiles (see also the appendix of D06).

The models are evolved for a maximum duration of 13 gigayears 
(i.e., $\approx$ a Hubble time, $t_H$) so that we can see the 
evolution of the core structure, and compare the evolving 
profiles to those of observed clusters.

\begin{figure}
\centering
\includegraphics[width=8.4cm]{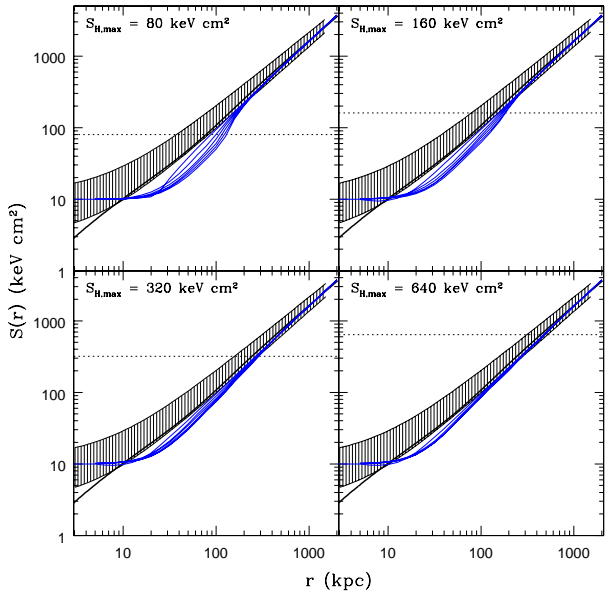}
\caption{Entropy profile predictions for the AGN ``wood stove'' 
heating model (with no preheating).  The solid black curve 
represents the pure cooling model for a cluster with a typical 
total mass of $M_{200} = 10^{15} M_\odot$.  The shaded region 
represents the observed profiles of D06, V06, and P06 (see Fig.\ 
1).  The solid blue curves represent the predictions of the 
``wood stove'' model.  The top curve shows the profile when the 
AGN turns on for the first time.  Each subsequent curve (top to 
bottom) shows the predicted profile after 1 gigayear of 
evolution, with the bottom most curve representing 5 gigayears 
following the initial burst.  The panels show the results for 
various choices of the maximum central heating level, $S_{\rm 
H,max}$.  The horizontal dotted lines, which indicate $S_{\rm 
H,max}$, can be used to estimate the maximum radius to which the 
bubbles can float convectively. In each case, $S_{\rm H,min} = 
10$ keV cm$^2$.
 }
\end{figure}

\subsubsection{Comparison to observed CC clusters}

Plotted in Figure 11 are the entropy profile predictions of the 
``wood stove'' model for various choices of the maximum 
heating level $S_{\rm H,max}$.  Since the clusters are in 
convective equilibrium, $S_{\rm H,max}$ can also be thought of 
as a parameter that controls the maximum radius that the bubbles 
can float to.  If the maximum heating level is relatively 
small (as in the top two panels of Fig.\ 11), such that the 
material is being circulated only within the central 100 kpc or 
so, we find that entropy profiles evolve relatively rapidly away 
from the observed profiles at radii of roughly $10$ kpc $< r < 
200$ kpc.  While the AGN prevents gas from dropping below the 
minimum at the centre, it clearly does not not prevent material 
further out in the cluster from cooling. Over the period of 13 
gigayears, this cooling material piles up near centre, driving 
the theoretical profiles away from the observational data.

If the maximum heating level is increased, some fraction of the 
heated material floats beyond the cooling radius.  In this case, 
the piling up of cool gas at the centre is reduced significantly.  
However, even in this case a satisfactory match to the mean 
observed profile is not achieved.  Instead, the ``wood stove'' 
heating simply maintains the pure cooling entropy profile 
(except, of course, near $S_{\rm H,min}$).  Physically, this is 
what one should expect, since the heating source has been 
designed to affect only the gas at the very centre.

Therefore, it appears that central heating by itself cannot 
solve the problem.  
Mathews and collaborators have argued that such a model {\it 
can} maintain the observed profiles of CC systems for long time 
scales.  However, their initial conditions are based on the 
{\it observed} properties of prototypical CC clusters, which 
can be maintained for long times with this simple heating model.  
Our results show that this model is not sufficient if one starts 
from initial conditions based on radiative cooling applied to 
clusters with properties guided by cosmological simulations.  

\subsection{Preheating + AGN ``wood stove'' heating}

What is needed is a mechanism for starting the ICM off on a 
higher adiabat, which the AGN ``wood stove'' can then maintain.  
Naively, one way to do this is to simply lower the initial 
baryon fraction, $f_b$, of clusters by assuming a larger value 
of $\Omega_m$ or a smaller value of $\Omega_b$.  (Although the 
observed baryon fractions of clusters do not permit large 
adjustments of these parameters; see McCarthy et al.\ 2007b.)
However, this would only have the effect of shifting the 
theoretical profiles in Fig.\ 11 up or down. What is really 
needed is a change in shape of the profiles by preferentially 
raising the entropy at small {\it and} intermediate radii.  This
is exactly what preheating achieves.  

Of course, in our simplified model neither preheating nor 
convectively rising gas provide heat to the gas at 
intermediate/large radii (following cluster formation, that 
is), so the tendency of a model that combines preheating and 
AGN central heating is to evolve towards the pure cooling 
state.  However, as we demonstrate below, the timescale for 
this to occur is long compared to the age of the cluster if 
the level of preheating is sufficiently high.  For this 
reason, the scenario is worth a closer look.  Moreover, since 
we have already seen that preheating is required to explain 
systems without cool cores (i.e., the NCCs), it is only natural 
to apply the same model to CC systems as well.

\subsubsection{Applying preheating + AGN ``wood stove'' heating 
to CC clusters}

To mimic the effects of preheating we modify the baseline 
gravitational entropy profile (eqn.\ 3) by adding a pedestal of 
amplitude $S_{\rm PH}$.   The model is then evolved with 
radiative cooling and AGN heating as described in the previous
section. We evolve the model for a maximum duration of $13$ 
gigayears.  We will see that relatively small differences in 
the initial level of heating lead to a very large difference in 
entropy at the final time.

\begin{figure*}
\centering
\includegraphics[width=13.5cm]{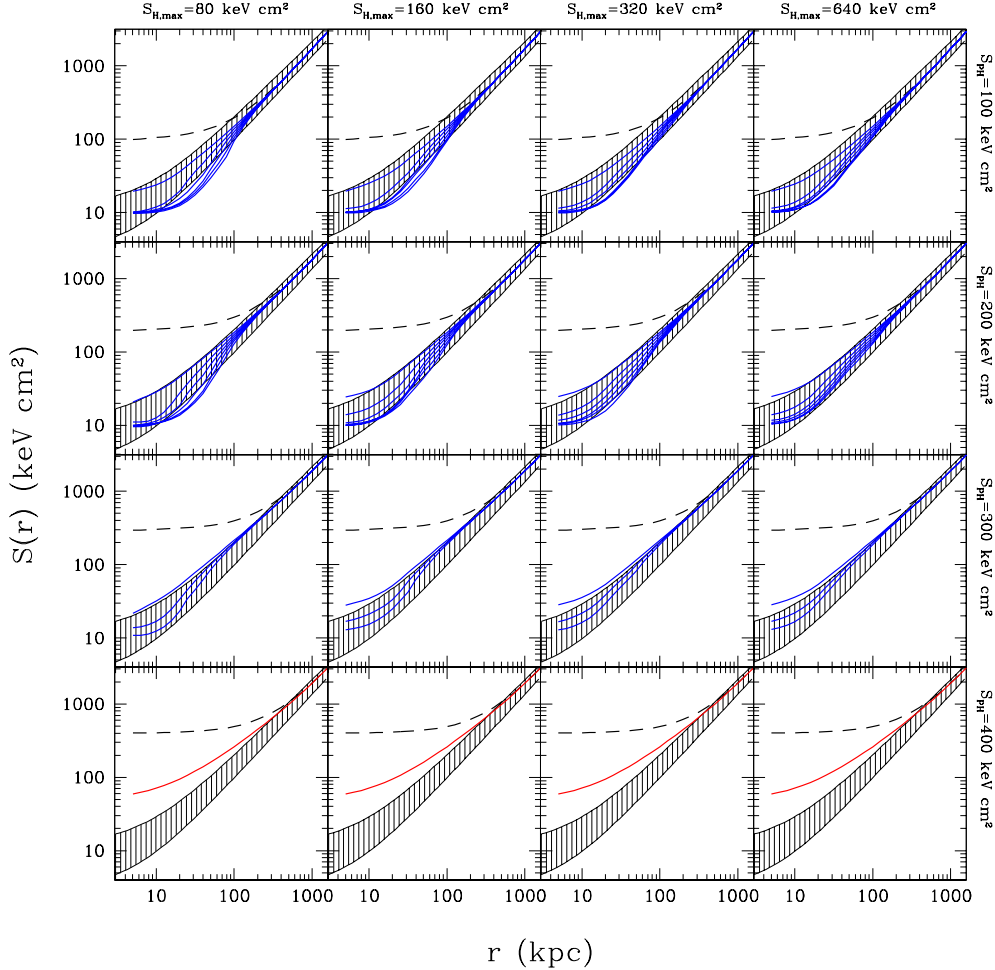}
\caption{Entropy profile predictions for the preheated AGN 
``wood stove'' heating model.  In each panel the shaded 
region represents the observed profiles (see Fig.\ 1) and the 
dashed line represents the initial profile of the model.  In the 
top three rows, which represent preheating levels of $S_{\rm PH} 
= 100$, $200$, and $300$ keV cm$^2$, the solid blue curves 
represent the predictions of the model.  In these cases, the top 
blue curve shows the profile when the AGN turns on for the first 
time.  All subsequent curves (top to bottom) show the evolved 
profile in intervals of 1 gigayear following AGN ignition, with 
the bottom curve representing 5 gigayears after the initial 
burst.  In the bottom row, corresponding to $S_{\rm PH} = 400$ 
keV cm$^2$, there is insufficient time for  catastrophic cooling 
to occur and, consequently, no heating is initiated.  The solid 
red line in these panels represents the final profile, after 13 
gigayears of cooling.
 }
\end{figure*}

We present the results in Figure 12 for various choices of the 
level of preheating.  As the top panels indicate, preheating the 
gas to a relatively mild level of $100$ keV cm$^2$ certainly 
yields an improvement relative to the no-preheating case, at 
least initially.  However, after a few gigayears of evolution the 
theoretical profiles tend to evolve to the lower envelope of the 
data.  This conclusion holds almost irrespective of the 
``wood stove'' heating input.  Increasing the level of 
preheating to $200$ keV cm$^2$ or $300$ keV cm$^2$, however, 
results in entropy profiles that are in much better agreement 
with the observed profiles and remain so for many gigayears 
following AGN ignition.  Note that in the case of $S_{\rm 
PH}=300$ keV cm$^2$ (second row from the bottom), only three 
theoretical curves are  visible in each panel.  This is simply 
because it takes nearly 10 gigayears of evolution before the 
onset of central catastrophic cooling, leaving only 3 gigayears 
for the AGN ``wood stove'' to operate.  Thus, 300 keV cm$^2$ 
corresponds nearly to the cooling threshold for massive 
clusters (see also Voit et al.\ 2002; M04).  This is clearly 
illustrated by the bottom row of panels, which demonstrates 
that if the system is preheated to $S_{\rm PH}=400$ keV cm$^2$ 
there is insufficient time for catastrophic cooling to occur 
and, consequently, no heating is initiated.  Systems preheated 
to this degree (or higher) clearly cannot explain CC clusters 
but may provide a viable explanation for NCC systems (see \S 
4.2).

Based on Fig.\ 12, we conclude that combining preheating with 
the AGN ``wood stove'' model can successfully match the observed 
properties of CC clusters.  This conclusion is not particularly 
sensitive to the level of preheating or the strength of the 
heat input.  In terms of preheating, entropy injection at the 
level of $100$ keV cm$^2$ $< S_{\rm PH} < 300$ keV cm$^2$ 
results in good agreement with the CC clusters data.  If 
the level of preheating is much lower than this, the model 
basically reverts back to the no-preheating case (see 
Fig.\ 11).  If the level of preheating is much higher than this, 
then: (1) one could not match the observational data for CC 
clusters at intermediate radii; and, (2) there would be 
insufficient time by the present day for clusters to form cool 
cores.  

This last point is particularly interesting since both conditions 
did not have to be met simultaneously.  If, for example, the 
data at intermediate radii actually required heating levels of 
much greater than $300$ keV cm$^2$, this would actually rule out 
preheating since we do not expect cool cores to form by the 
present day when the ICM has been preheated at levels much 
greater than this (see the panels in the bottom row of Fig.\ 12).  
In this case, we would be forced to abandon preheating for a 
more complex form of highly-energetic distributed heating 
following cluster formation.  Instead, the entropy profiles of 
observed CC clusters can accommodate entropy injection only up 
to the level of $300$ keV cm$^2$ or so, which is consistent 
with the preheating scenario.

A cluster preheated to the lower bound of $\approx 100$ keV 
cm$^2$ requires $\sim 3$ gigayears of cooling to form a cool 
core.  For a concordance cosmology, this implies 
that there should be no CC clusters at $z > 2$.  In fact, the 
maximum redshift that one should see CC clusters to under this 
scenario would likely have to be somewhat lower than this, since 
$t_{\rm age} < t_{H}$.  Interestingly, Vikhlinin et al.\ (2007) 
have recently reported a deficit of CC clusters at high $z$ 
compared to their abundance today (see also Santos et al.\ 
2008).  They suggest that increased 
merger activity might account for this finding, but this 
phenomenon could also plausibly be accounted for by preheating.  
However, a quantitative comparison with the data is required 
before a more definitive statement can be made.  We leave this 
for future work.

While we have demonstrated that preheating combined with the
AGN ``wood stove'' model is successful at matching the mean radial 
structure of CC clusters, the model could potentially violate a number
of observational constraints. For example, it has been argued that 
the convective outflows triggered by significantly heating gas 
at the cluster centre would rapidly destroy the observed peaked 
metallicity gradients in many CC clusters. We consider these
constraints in Appendix~E, where we show that the model is entirely
consistent with current data. Equally importantly, the simplest 
``wood stove'' model neglects heating in the ambient medium 
associated with the inflation of bubbles as well as distributed 
heating by the bubbles as they rise.  We discuss these and 
other relevant issues in Appendix E.  We stress, however, that 
even if the simplest ``wood stove'' model could be demonstrated 
to be overly naive, this does not detract from our main 
conclusion that preheating is required to set the stage for 
present-day heating (whatever its form) in CC clusters.

\section{Summary \& Conclusions}

Although there has been much progress in understanding the 
present-day competition between heating and cooling in clusters 
in recent years, little is currently known about the early 
heating processes that set the properties of the ICM in the first 
place.  The observed spread in cluster core morphologies is 
another poorly understood feature of clusters and may provide 
an important clue to these early heating processes.  The 
purpose of the present study was to shed light on both of these 
important issues.  To quantify the effects of non-gravitational 
heating in clusters, we developed a physical model for the ICM 
that includes the effects of gravitational shock heating, which 
dominates at large radii, and radiative cooling, which is 
important at small radii.  We refer to this as the ``pure 
cooling model''.  

A comparison of the pure cooling model to the observations yields 
the following results:

\begin{itemize}
\item{At large radii, the profiles of CC and NCC clusters scale 
according to the gravitational self-similar model, but at 
small radii the profiles of both types of clusters clearly 
deviate from the predictions of the pure cooling model.}
\item{A simple modification of the predicted entropy 
profile of the pure cooling model (see eqn. 8) yields an 
excellent simultaneous match to the entropy and density profiles 
of CC clusters and to their luminosity-temperature relation and 
high resolution X-ray spectra as well (see \S 3.1).}
\item{NCC clusters, on the other hand, show a much broader 
scatter in entropy at small radii but their gas density profiles 
and luminosity-temperature relation can still be described well 
by a simple pedestal modification (of varying amplitude) to the 
pure cooling model (see \S 3.2).} 
\item{The energy required to drive most clusters to their observed 
distributions from the pure cooling state is huge (far exceeding 
the most energetic AGN bursts observed in the nearby universe for 
most clusters) if the heating occurs internally following cluster 
formation (see \S 4.1).  Alternatively, if clusters were 
preheated at $1 < z < 3$ or so, the required energetics can be 
reduced by up to two orders of magnitude (see \S 4.2).  This puts 
even the most extreme NCC clusters within reach of observed AGN 
power output.}
\item{The short central cooling times of CC clusters imply that, 
in addition to preheating, there must be significant present-day 
heating as well.  (NCC clusters, on the other hand, require no 
further heating.)  In \S 5, we showed using a simple model 
that combining preheating and a AGN feedback loop, which is 
triggered at the onset of catastrophic cooling, can successfully 
explain the observed properties of CC clusters without fine 
tuning of the preheating or AGN feedback strengths.  We note, 
however, that AGN central heating by itself (i.e., without 
preheating) cannot explain the observed systems.
}  
\end{itemize}

Based on the above findings (and those of our previous studies), 
we envisage a scenario in which {\it all} clusters are preheated 
to varying degrees.  The inferred spread in the central 
entropies of present-day clusters is large, ranging from $\sim 
10$ keV cm$^2$ up to possibly $700$ keV cm$^2$.  How would such a 
large spread be achieved?  In particular, since clusters are 
built hierarchically from many smaller systems, the implication 
is that NCC clusters would somehow have been built from systems 
that preferentially had higher levels of preheating activity 
(associated with galaxy- and supermassive black hole formation).  
As a consequence, one therefore might expect NCC clusters to have 
larger galaxy/bulge populations.  This is presently an open 
question.  However, it is important to recognise that the 
spread in present-day central entropies is {\it not} equivalent 
to the spread in intrinsic preheating levels.  Radiative cooling 
losses following preheating act to amplify the intrinsic 
spread.  For example, we have demonstrated that for systems 
preheated to less than about $\sim 300$ keV cm$^2$, 
there is sufficient time for cooling to lower the central 
entropy to the level observed in CC clusters.  On the other 
hand, if the level of preheating is as high as, say, $600-700$ 
keV cm$^2$, the effects of cooling will be minimal following 
preheating.  In this way, a factor of only two in the intrinsic 
spread of preheating levels can give rise to almost two orders 
of magnitude in dispersion in the distribution of present-day 
central entropies.  

An additional way to generate scatter in the central entropies 
of local clusters is via cluster mergers.  While such mergers do 
not seem capable of transforming a CC cluster into a NCC cluster 
(see \S 1), if the core has already been heated (say, to a 
moderate CC state) the impact of the merger could be significant.  
This is simply because the entropy generated in the shock 
depends on the inverse of the pre-shock density to the 
two-thirds power (e.g., Voit et al.\ 2003).  Heating the gas 
non-gravitationally first lowers the density of the ICM which, in 
turn, will increase the efficiency of the shock heating that 
occurs as a result of the merger (e.g., Borgani et al.\ 2005; 
McCarthy et al., in preparation).

Whatever the mechanism that generates the dispersion, we 
hypothesise that the systems that were significantly preheated 
may have evolved into NCC clusters we see today.  Systems that 
were initially heated to $\sim 300$ keV cm$^2$ or lower, 
however, have sufficient time to develop cool cores.  In this 
case, eventually catastrophic cooling feeds a supermassive 
black hole at the cluster centre which initialises a jet that 
heats the gas and inflates buoyantly rising bubbles.
However, the signature of preheating is still left behind in 
these systems as an excess entropy (relative to the pure 
cooling profile) at intermediate radii ($r \sim r_{\rm cool}$).

Our findings do not rule out the contribution of additional 
sources of heating such as thermal conduction or distributed AGN 
heating.  However, we would argue that our model represents the 
most physically simple and energy efficient heating scenario yet 
proposed that is consistent with the observed properties of the 
whole (i.e., both CC and NCC) cluster population.  Detailed high 
quality measurements of large {\it representative} samples would 
go a long way towards confirming or ruling out our proposed 
picture.  Fortunately, we may not have to wait long, as the 
release of several such samples appears imminent (see, e.g., 
Sanderson et al.\ 2006; Pratt et al.\ 2007; Vikhlinin et al.\ 
2007; Hudson, Reiprich et al.\ in prep.).

\section*{Acknowledgments}

The authors thank the referee for suggestions that significantly 
improved the paper.  They also thank Brian McNamara, Alastair 
Edge, and Mark Voit for useful comments and Megan Donahue, Don 
Horner, Gabriel Pratt, Alexey Vikhlinin, Joe Mohr, and Alastair 
Sanderson for providing their observational data.  IGM thanks 
Mark Fardal and Andisheh Mahdavi for helpful discussions and 
acknowledges support from a NSERC Postdoctoral Fellowship.  AB 
and MLB acknowledge support from NSERC Discovery Grants.  AB 
also extends his appreciation to the Universities of Durham and 
Oxford for hosting him during his tenure as the Leverhulme 
Visiting Professor, during the course of which the present 
study was carried out. RGB acknowledges the support of a PPARC 
senior fellowship.

\section*{Appendix A: ACCURACY OF THE PURE COOLING MODEL}

The key approximations are:
\\
\\
\noindent I. {\bf Quasi-hydrostatic equilibrium.}  At all times 
the ICM is assumed to be in HSE within the potential 
well of the cluster.  This assumption is motivated by the 
results of 
cosmological simulations (e.g., Evrard et al.\ 1996; Lewis et 
al.\ 2000; Kravtsov et al.\ 2005; Rasia et al.\ 2005).  While 
such simulations do show 
deviations from HSE (e.g., due to merger events), they are 
typically only at the level of 10\%.  This result is re-enforced 
by high resolution simulations of idealised cluster mergers 
(e.g., Poole et al.\ 2006).  Depending on the depth of 
the potential well and the properties of the ICM, 
quasi-HSE can also potentially be violated in the central 
regions by simple inflow from loss of pressure support due to 
radiative cooling.  However, for the massive systems of interest 
here, we verify that, at all radii greater than a few kpc, 
$t_{\rm cool}(r)$ exceeds the local free-fall time and, 
therefore, the ICM can be considered to be in quasi-HSE.
\\
\\
\noindent II. {\bf Static gravitational potential.}  We assume the 
cluster gravitational potential does not evolve with time.  This 
will be violated if the cluster grows significantly via mergers 
and accretion or if large quantities of baryons are able to 
completely cool and subsequently settle at the cluster centre.  
With regards to the former, cosmological simulations indicate 
that most of the mass of a typical rich cluster has been in 
place since $z > 0.5$ (e.g., Cohn \& White 2005).  Thus, for a 
concordance cosmology the typical age of a cluster corresponds 
to $t_{\rm age} > 5$ gigayears, which exceeds its dynamical time 
by a factor of a few, implying the equilibrium assumption is 
justified.  As for the latter, if a significant fraction of 
the baryons are able to cool and collect at the cluster centre, 
this can significantly deepen the gravitational potential there 
(see, e.g., Lewis et al.\ 2000).  However, it is now well known 
that large quantities of cold baryons are {\it not} piling up in 
the centres of clusters (e.g., Balogh et al.\ 2001; Peterson et 
al.\ 2003).  The implication is that some form of  
non-gravitational heating is offsetting radiative cooling.  We 
therefore neglect the effects of cold gas and stars on the 
gravitational potential, assuming by default that some source of 
non-gravitational heating (significantly moderates or even wins 
over radiative cooling) is initiated shortly following the onset 
of catastrophic cooling.
\\
\\
\noindent III. {\bf Spherical symmetry.}  Our theoretical 
clusters are spherically symmetric. With suitable redefinition
of the coordinate system, the model can also be applied to 
ellipsoidal systems. Of course, real systems deviate from such 
strict symmetry.  One can minimise this  bias by comparing to 
large samples of clusters, as we do in \S 3, so that any 
asymmetries present are averaged out.  This bias can be 
minimised further by selecting observed systems that appear 
visually relaxed (see, e.g., Poole et al.\ 2006).  The cluster 
samples studied in \S 3, particularly in \S 3.1, have largely 
been constructed so that the clusters meet some isophotal 
regularity criterion.
\\
\\
\noindent IV. {\bf Single-phase cooling.}  Our models 
assume that at each radius the properties of the gas can be 
characterised by a unique entropy, density, temperature, etc.  
Consequently, if the ICM is actually a multi-phase medium a 
comparison of these models to the observational data could lead 
to spurious conclusions.  Over the past few decades there has 
been debate about the role of multi-phase cooling in clusters.  
We examine this issue in more detail in Appendix B and 
demonstrate that the observational evidence usually put forward 
in support of multi-phase cooling is actually fully consistent 
with our single-phase models.  We argue further that the current 
suite of multi-wavelength cluster observations strongly favours 
single-phase cooling for the vast bulk of the gas in massive 
clusters.
\\
\\
Finally, we have tested the numerical accuracy of our 
cooling algorithm against the analytic cooling wave similarity 
solutions of Bertschinger (1989) and find excellent accordance.

\section*{Appendix B: SINGLE- VS. MULTI-PHASE COOLING}

There has been debate over the role of multi-phase cooling in 
the ICM for a couple of decades now.  It is beyond the scope of 
this paper to summarise in detail all of the many facets of this 
discussion.  For ease of discussion, therefore, we summarise 
just the main arguments usually put forward in support of a 
multi-phase ICM and comment on them in the context of our 
single-phase models.

There are three main arguments that have traditionally been used in 
support of a multi-phase ICM: (1) multi-temperature plasma models 
give better fits to the observed X-ray spectra of cool 
core clusters; (2) the observed X-ray surface brightness/ICM gas 
density profiles are too steep compared with a single-phase 
cooling model; and (3) the ICM is thermally unstable.  The first 
two arguments are observational in nature while the last one is 
theoretical.  We address each of these arguments in order.

Multi-temperature plasma models have been demonstrated to 
yield statistically better fits than single-temperature models
to the X-ray spectra of cool core systems (e.g., Allen 2000), even 
when the spectra are properly deprojected (Kaastra et al.\ 2004).  
But clearly an important consideration is the width of the annulus 
from which the X-ray spectra are extracted.  If the width is 
larger than the typical scale over which the temperature profile 
changes appreciably, one should {\it expect} a multi-phase model to 
fit the spectrum better than a single-phase model even if the gas 
is single phase at all radii within that annulus.  Normally it is 
the width of the PSF X-ray instrument that dictates the minimum 
width one will select for each annulus.  For example, {\it 
ASCA} has a broad PSF with a FWHM of $\approx 3$ arcminutes, which 
corresponds to a physical size of $\approx 170$ kpc for a cluster 
at $z = 0.05$.  This is comparable or slightly larger than the 
cooling radius of a typical cool core system, so it is not 
surprising that multi-temperature plasma models fit {\it ASCA} 
spectra better than single-temperature models.  The {\it Chandra} 
X-ray Observatory, with its good spectral resolution, wide 
passband, and, importantly, its high spatial resolution ($\approx 
1$ arcsecond) should place the best constraints on whether the 
ICM is actually multi-phase.  Numerous recent studies (e.g., David 
et al.\ 2001; V06; D06) have demonstrated that, probably with the 
exception of the innermost radial bin, the spatially-resolved 
{\it Chandra} spectra of cool core clusters are fit well by
single-temperature models.  Multi-temperature models also fit the 
data well, but the improvement over the single-temperature models 
is not statistically significant.  Therefore, with the exception 
of the innermost radial bin (which for nearby systems typically 
corresponds to a physical size of a few up to 10 kpc or so), there 
is no longer strong spectral evidence for multi-phase cooling in 
cool core clusters.  Whether or not the gas at very small radii 
is cooling in a multi-phase fashion will require even higher 
resolution observations that can rule out the possibility that the 
spectra can be simply explained with a single-phase plasma 
characterised by a steep temperature gradient.

The observed surface brightness/gas density profiles of cool core 
clusters also potentially yield important clues about the nature of 
cooling of the ICM.  Several studies (e.g., Nulsen 1986; Johnstone 
et al.\ 1992; Peterson \& Fabian 2006) have argued that the 
observed profiles are not steep enough to be explained by a 
single-phase cooling model.  The plots in \S 3.1.3 confirm this 
is true when a comparison is made between the pure cooling 
model\footnote{We note that our pure cooling model is the 
configuration of the ICM at the onset of catastrophic cooling.  
If one instead evolved this model further until a steady-state 
cooling flow developed, as assumed by Nulsen (1986), Johnstone 
et al.\ (1992), and Peterson \& Fabian (2006), the resulting 
gas density profiles would be even more strongly peaked.}
and the observed profiles.  This gas density problem is often
recast in terms of a (so-called morphological) ``mass deposition 
rate problem''.  The argument is that single-phase models 
supposedly predict $\dot{M}$ is constant with radius inside the 
cooling radius, whereas the observations typically show that 
integrated mass deposition rate scales nearly linearly with 
radius $\dot{M} \propto r^{\sim 1}$.  Nulsen (1986) has shown 
that this kind of behaviour can be broadly accounted for by a 
model in which the ICM cools in an inhomogeneous multi-phase 
fashion.  As such, the fact that observed cool core systems have 
relatively ``shallow'' surface brightness profiles is often 
taken as support for a multi-phase ICM.  However, the above 
argument against a single-phase ICM is flawed because it 
neglects the effects of non-gravitational heating.  As we have 
shown throughout the paper, a simple modification of the pure 
cooling entropy profile results in excellent simultaneous 
agreement with a wide variety data.

Finally, regarding thermal instability, in the absence of any 
sources of non-gravitational heating the ICM can easily 
be demonstrated to be thermally unstable by considering the 
generalised Field criterion (e.g., Balbus 1986).  If the 
perturbations are initially small, the growth time is two-thirds 
the local isobaric cooling time (see Kim \& Narayan 2003).  
However, in a quasi-hydrostatic cooling medium, the cooling time 
of the gas is always close to the inflow time (see, e.g., 
Bertschinger 1989).  
(We have verified this for our algorithm by tracking individual 
gas shells in our 1-D model as they flow from the cooling radius 
to the cluster centre.)  Since the growth time of perturbations is 
only slightly faster than the cooling time, this implies the 
growth time and the inflow time are also comparable.  So gas that 
is initially perturbed at some radius (say, the cooling radius) 
only becomes non-linear as it reaches the centre of cluster.  
Following this line of argument, B89 have estimated, for example, 
that a 10\% density perturbation at the cooling radius in M87 
would only become nonlinear at $r \approx 15$ kpc.  However, if 
the density perturbations are much larger than this (i.e., 
non-linear, or nearly so, to begin with), as in the models of 
Nulsen (1986), a multi-phase medium can potentially result.

Recent observational results suggest that any multi-phase cooling 
present in clusters is limited in scope.  First, as noted in \S 
3.1.5, {\it XMM-Newton} spectra indicate that very little gas is 
cooling below approximately one half the ambient ($\approx$ 
virial) temperature.  Assuming that the perturbations are in 
pressure equilibrium with 
the ambient ICM, this implies that the variation in density of 
the ICM at a fixed location is at most a factor of two.  But in 
order to retain consistency with the fact that spatially-resolved
{\it Chandra} spectra are fit well by single-temperature models, 
the variation in density will likely have to be much smaller than 
this.  Furthermore, recent surveys for cold gas and dust in cool 
core clusters (e.g., Edge et al.\ 2002; Edge \& Frayer 2003; 
Egami et al.\ 2006) generally indicate that, if present, they 
are typically confined to small radii ($r \approx 20$ kpc or 
so).  The implication is that either the typical ICM density 
perturbation is quite small (which could account nicely for the 
limited spatial extent of cold gas and dust) or that some form of 
non-gravitational heating is counteracting the growth of 
perturbations (or that both are true).  In any case, the above 
argues in favour of using single-phase models to study the vast 
bulk of the ICM in galaxy clusters.

\section*{APPENDIX C: ENERGETICS OF CLUSTER HEATING}

At first sight it seems strange that the energy required to 
modify the gas distribution in a cluster should depend on 
whether the energy is injected before or after cluster collapse. 
Below we outline the thermodynamics of the heating process and 
explain why greater energy is required after the collapse of the 
cluster.

We should not be surprised that the two heating scheme require
different energies to achieve the same change in gas entropy.
This is what powers your car.
Consider two isolated parcels of gas.  We will heat one, keeping 
its density (and hence volume) fixed.  We will adiabatically 
compress the other, and then heat it at constant density such 
that the two parcels of gas have the same entropy.  We then let 
this second parcel expand back to its initial density.
The two parcels have the same 
initial and final states but require different heat inputs, as 
we will show below. Since the entropy is conserved during the 
expansion and compression, the entropy change of the two 
parcels is the same. However, since the entropy change is 
directly proportional to the (logarithm of the) change in 
thermal energy divided by its density to the two-thirds power, 
it is immediately apparent that the amount of energy that must 
be thermalised in the gas to raise its entropy by some 
specified amount depends on the density of the gas at the time of 
heating and that it is different for the two parcels.  In 
particular, the higher the gas density the more energy that is 
required to reach the desired entropic state.  Applying this 
simple rationale to galaxy clusters (as done, e.g, by 
Ponman et al.\ 1999; Balogh et al.\ 1999; Tozzi et al.\ 2000; 
Tozzi \& Norman 2001), one concludes that in order to achieve 
the same ICM state (e.g., the observed state at $z \sim 0$), it 
is energetically more efficient to heat the ICM at high 
redshift prior to  collapse (i.e., ``preheat''), when its mean 
density was lower, than it is to heat it today after the 
cluster has fully collapsed and virialised.

An obvious question is ``what happens to the extra energy?''. 
In the case of an isolated parcel of gas, the additional heat
energy is balanced by the $PdV$ work required to compress the 
system and then bring it back to the initial density.  
The heat input is given by 
$$ \Delta Q = {3\over2} V_h^{-2/3} (K_f - K_I)$$
where $V_h$ is the volume of the gas when the heat input occurs,
$K_i$ and $K_f$ are the initial and final adiabats of the gas. 

Demonstrating that the simple rationale of the isolated gas 
parcel applies in detail to the case of galaxy clusters where 
the gas is gravitationally bound to a massive (evolving) dark 
matter potential is slightly more involved but is still 
tractable.  As we show below, the difference in the energies 
between the preheating and internal heating cases corresponds 
to the work that must be done by the gas (in the internal 
heating case) against the gravitational potential and on the 
external medium as it expands to the same final configuration. 
To demonstrate this, we make an explicit comparison of the 
required internal and pre-heating energetics for a $M_{200} = 
10^{15} M_\odot$ cluster with a final entropy profile that is 
given by the addition of a $500$ keV cm$^2$ pedestal to the 
pure cooling profile.

Let us first consider the case where the heating takes place 
after the cluster has collapsed (i.e., internal heating).
Key to a good estimate of the energetics for this form of 
heating is an accurate estimation of the surface bounding 
pressure term (as noted in \S 2, the properties of the ICM are 
set by its entropy distribution, the depth of its dark matter 
potential well, and the bounding pressure at the virial 
radius).  The bounding surface pressure term is estimated as 
follows.  We place the pure cooling entropy profile in 
HSE within a NFW dark matter potential.  The bounding surface 
pressure is whatever it needs to be so that the total mass of 
gas within $r_{200}$ matches that in the simulations (see 
eqn.\ 6).  In the case of a typical $M_{200} = 10^{15} 
M_\odot$ cluster, we find the bounding pressure is $\sim 
3\times 10^{-13}$ ergs/cm$^3$ at $r_{200}$. (A virtually 
identical result is obtained if one instead uses the baseline 
gravitational entropy profile.)

In order to heat the gas to the desired state, internal 
energy must be added to the system.  We find that the total 
energy required to modify the Lagrangian entropy distribution of 
the pure cooling model so that it matches the desired final 
Lagrangian entropy distribution of the $500$ keV cm$^2$ pedestal 
model is $\approx 3.8\times10^{63}$ ergs. Heating the gas 
causes it to expand, so that the gas mass within $r_{200}$ of 
the final system is only about 70\% of what it was before 
heating (this number is sensitive to the bounding surface 
pressure, which is why we required an accurate estimate of this 
quantity).  So the estimate of the energy required to heat the 
gas given above is only for the gas that remains within 
$r_{200}$ of the final system.

As the gas expands, much of the internal energy added to the 
system will be used to do work against the dark 
matter-dominated gravitational potential of the cluster (i.e., 
as the gas expands, work is required to lift gas sitting at 
larger radii).  We calculate the work done as follows.  For 
each gas mass shell, we calculate the volume occupied by the 
shell before, $V_i = 4 \pi \Delta r_i^3/3$, and following, 
$V_f = 4 \pi \Delta r_f^3/3$, the heating (where $\Delta r_i$ 
and $\Delta r_f$ are the initial and final thicknesses of the 
shell).  The net work done by a shell against the potential is 
just the integral of $P(V) dV$ over the interval $V_i$ to 
$V_f$.  The initial pressure, $P(V_i)$, is the pressure of the 
shell at the instant it has been heated (but before expansion), 
while $P(V_f)$ is the pressure of the shell in the final 
(hydrostatic) configuration.  The expansion of the shell in 
going from $V_i$ to $V_f$ is adiabatic, such that $P \propto 
V^{-5/3}$.  Summing up the work done by all of the shells 
yields the total work done against the potential.  
For the specific example, we find that heat input required is 
$3.8\times10^{63}\erg$.  As the gas expands and rises in the 
potential, the total $PdV$ work done during expansion is 
$3.8\times10^{63}\erg$. The majority of this energy goes into 
lifting the gas in the gravitational potential  
($1.7\times10^{63}\erg$) while the remaining energy goes into 
$PdV$ work done at the outer boundary (lifting the exterior gas 
in the  potential). The change in the internal of the system 
is much smaller than the above quantities, only $\sim 
10^{61}\erg$. All these quantities refer to the gas that 
remains within $r_{200}$.
Thus we find that 99.9\% of the energy added to the system goes 
into work done against the potential, while the remaining 
energy stays in the form of internal energy.  

Let us now consider the preheating case.  To a good 
approximation, the addition of a pedestal of $500$ keV cm$^2$ 
to the pure cooling model can be achieved by uniformly 
preheating the proto-ICM to this level (see, e.g., Balogh et 
al.\ 1999; Voit et al.\ 2003; Lu \& Mo 2007; Younger \& Bryan 
2007).  As the density of the proto-ICM is much lower than the 
density typical of a collapsed cluster, the required amount of 
energy that must be added in the preheating case (see eqn.\ 
11) is much lower than in the internal heating case. 
For example, if we heat the gas when it has a mean density 
corresponding to $z=1$, the energy required is  $2.6 \times 
10^{61}\erg$.
Note that while some of this energy will be converted into 
work done in the expansion of the gas, this energy will 
eventually be retrieved when the system turns around and 
collapses (i.e., infall kinetic energy of the shells will be 
converted back into internal energy via shock heating).

To complete our energy budget, we need to consider the energy
that is thermalised during the system collapse. Without 
pre-heating, the system collapses to a denser, more bound 
configuration: yet we have seen that the internal energies of 
the two systems (measured within the Lagrangian region) are 
quite similar. While this is puzzling at first sight, the key 
difference is that initially the gas collapses along with the 
dark matter into a steepening gravitational potential, while
when it is heated the potential is fixed. During collapse, the 
gas is shock heated and then compressed as the rest of the 
cluster collapses around it (for a detailed discussion, see 
Bertschinger 1989, Abadi et al.\ 2000, Voit et al.\ 2003). To 
reverse the collapse and reach the desired final state, we must 
heat the gas, puffing it out by doing $PdV$ work against a fixed 
(and deep) dark matter potential. A detailed calculation of the 
shock heating and $PdV$ work during collapse is beyond 
the scope of this appendix, 
but we can estimate an upper limit to the amplitude of this 
correction by assuming that the gas is compressed along a fixed 
adiabat. This corresponds to the case we presented in the 
initial discussion of compression in a gas cylinder, and 
amounts to ignoring the growth of the potential with time and 
the role of shocks in heating the gas. We are also ignoring the 
greater efficiency of shocks and $PdV$ work in the collapse of 
the 
preheated gas. Following this methodology sets an upper limit 
on the work done on the gas during the collapse of 30\% of that 
done by the gas during expansion. Thus (if the argument based 
on entropy is set aside) our comparison of the preheating and 
in-situ heating energies could be modified by up to 30\% to 
account for the work done in the collapse. Nevertheless, the
central result is still clear: ``preheating'' is approximately 
two orders of magnitude more efficient than heating the gas halo 
once it has formed.
 
Finally, we stress that while we concluded in \S 4.1 that 
radiative cooling losses were not very important in the 
internal heating case {\it after the heating had taken place}, 
the amount of energy that must be radiated in going from the 
baseline gravitational model to the pure cooling model (in 
order to trigger the central black hole into action) is $\sim 
10^{62}$ ergs.  This alone is approximately an order of 
magnitude more energy than is required to preheat the gas.

\section*{APPENDIX D: THE ROLE OF THERMAL CONDUCTION}

The idea that thermal conduction could be 
important in offsetting radiative losses in CC clusters has been 
revived in recent years.  This is largely due to recent 
theoretical work (e.g., Malyshkin \& Kulsrud 2001; Narayan \& 
Medvedev 2001) that suggests that the magnetic fields present in 
clusters may not suppress conduction as much as previously 
thought.  If the magnetic field lines are chaotically tangled, 
the effective conductivity, $\kappa$, could be as high as 
0.3 times the unmagnetised Spitzer rate ($\kappa_{\rm s}$).  
As demonstrated by several different groups (e.g., Narayan \& 
Medvedev 2001; Zakamska \& Narayan 2003; Voigt \& Fabian 2004; 
Hoeft \& Br{\"u}ggen 2004), such minor suppression could allow 
conduction to balance radiative losses in the central 
regions of some, but not all, CC clusters.  (Although to 
balance radiative losses in detail seems to require that the 
suppression coefficient vary somewhat strongly as a function of 
radius, changing by up to factors of several within $r_{\rm 
cool}$; see Pope et al.\ 2006.)  But whether or not $f_{\rm s}$ 
is actually of order 0.3 is still not known.  Observations of 
sharp temperature discontinuities in many clusters provide some 
evidence that the suppression coefficient is actually much 
smaller than this (e.g., Markevitch et al.\ 2003).  Furthermore, 
theoretical work by Loeb (2002) suggests that if thermal 
conduction was not heavily suppressed, this would result in a 
significant fraction of the thermal energy of the ICM being 
transported out 
of the cluster interior and into the surrounding intergalactic 
medium.  The net result is that clusters would be much cooler 
than we observe them to be.

\begin{figure}
\centering
\includegraphics[width=8.4cm]{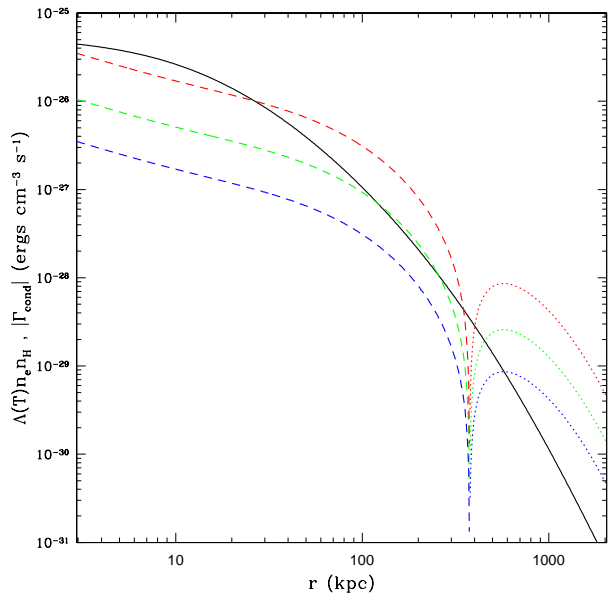}
\caption{Radiate cooling and conductive heating and cooling rates 
(per unit volume) for a typical $M_{200} = 10^{15} M_\odot$ 
cluster.  The solid black curve corresponds to the radiative 
cooling rate profile, whereas the blue, green, and red 
dashed/dotted curves represent the conduction rate profile for 
$f_{\rm s}$ = 0.1, 0.3, 1.0.  The dashed portion of these curves 
represents conductive heating while the dotted portion represents 
conductive cooling.
 }
\end{figure}

We leave these issues aside and simply ask how conduction 
characterised by a rate of $f_{\rm s} \sim 0.3$ would change our 
results or conclusions.  As noted above, conduction at this 
level is a non-negligible source of energy transport.  However, 
it has already been demonstrated (e.g., Zakamska \& Narayan 
2003; Voigt et al.\ 2004) that conduction at this level cannot 
{\it by itself} balance cooling in the most extreme cool core 
clusters or for moderate to low mass systems characterised by 
temperatures of $T_{\rm spec} < 4$ keV or so (note that the 
conductivity scales steeply with temperature as $\kappa \propto 
T^{5/2}$).  Although nature may not be so kind, we seek a single 
heating model to explain the properties of all clusters.  With 
this as our goal, we are interested in whether thermal conduction 
(with $f_{\rm s} \sim 0.3$) plus AGN ``wood stove'' heating can 
yield a 
satisfactory replacement to the preheating plus AGN heating 
model explored in \S 5.2.  For this to be the case, conduction 
must 
raise the adiabat of the gas at intermediate radii in the same 
fashion that preheating would.

In Fig.\ 13, we plot the radiative cooling and thermal conduction 
heating/cooling rates (per unit volume) for a typical $M_{200} = 
10^{15} M_\odot$ cluster set up according to the baseline 
gravitational model (\S 2.1).  The thermal conduction rate, 
$\Gamma_{\rm cond}$, is defined as:

\begin{equation}
\Gamma_{\rm cond} \equiv \frac{1}{r^2}\frac{d}{dr}(r^2 F_{\rm 
cond})
\end{equation}

\noindent where
\begin{equation}
F_{\rm cond} = -f_{\rm s} \kappa_{\rm s} \frac{dT}{dr}
\end{equation}

\noindent and $\kappa_{\rm s} \propto T^{5/2}$ and is defined in 
Spitzer \& Harm (1953).

At very small radii, thermal conduction with $f_{\rm s} \sim 
0.3$ cannot balance radiative losses, as expected.  However, this 
can plausibly be taken care of by AGN heating.  Of 
more 
interest is the conductive heating rate at intermediate radii.  
Interestingly, over the range 60 kpc $< r <$ 200 kpc (roughly), 
thermal conduction with $f_{\rm s} \sim 0.3$ can balance or even 
win over radiative losses.  This implies the entropy of the gas 
at these radii will be elevated by conduction, as required.  
Naively, therefore, this suggests that thermal conduction and 
AGN central heating in tandem has the correct behaviour to 
explain the observed structure of CC clusters.

However, there are at least two problems with this scenario.  
First, at large radii ($r > 400$ kpc or so) it is apparent that 
conduction substantially {\it increases} the cooling rate of the 
gas.  The reason for this behaviour is that there is a negative 
temperature gradient at large radii (this is a robust prediction 
of cosmological simulations that has been confirmed by  
observations) and, therefore, conduction acts to transfer energy 
down the gradient (see also Hoeft \& Br{\"u}ggen 2004).  We have 
investigated the seriousness of this effect by including 
conduction in our time-dependent 1-D algorithm.  Over the course 
of a Hubble time, we find that at large radii ($r \approx 
500-800$ kpc) conduction reduces the normalisation of the ICM 
entropy profile by at least $30\%$.  This is sufficient to yield 
serious tension with the amplitude of the observed profiles at 
large radii.  However, this issue could be mitigated to some 
degree by appealing to the fact that conduction has likely only 
been energetically important since the cluster obtained the bulk 
of its mass ($z < 0.5-1$ or so).

Secondly, thermal conduction by itself or combined with 
AGN ``wood stove'' heating has no hope of explaining NCC 
clusters.  
These 
systems seem to require large entropy cores ranging anywhere from 
100 keV cm$^2$ to 700 keV cm$^2$ (see Fig.\ 7).  However, we find 
that if conduction is extremely efficient ($f_s \sim 1$), such 
that it wins over radiative cooling at small radii, the largest 
entropy core it can produce is $\approx$ 100 keV 
cm$^2$ for a $M_{200} = 10^{15} M_\odot$ cluster (see also Hoeft 
\& Br{\"u}ggen 2004).  It cannot produce cores larger than this 
since this would require a negative temperature gradient at 
small radii.  The tendency of conduction, on the other hand, is 
to transfer heat to the central regions in order to establish an 
isothermal core.  Once an isothermal core is established the 
conductive heating rate drops to zero.

One of our main goals in this paper is to see if there is 
single heating model that can plausibly account for the observed 
properties of the whole cluster population (i.e., both CC and 
NCC systems).  This does not appear to be possible with thermal 
conduction or thermal conduction plus AGN central heating.  One 
needs to invoke some additional form of heating to rectify the 
issues mentioned above.  The preheating plus AGN 
central heating 
model presented in \S 5.2, however, seems capable of matching 
the properties of the observed systems and physically is more 
simple.  For this reason, it is our preferred model.

\section*{APPENDIX E: Other challenges for the ``wood stove'' 
model}

\subsection*{E.1 The effect on metallicity gradients}

An important consideration for the preheated AGN ``wood stove''
model is that of observed metallicity gradients in CC 
clusters.  It has been argued that the convective outflows that 
result from strong central heating would rapidly destroy the 
gradients.  However, the metallicity peaks 
in CC clusters are typically quite wide, with $\Delta r \approx 
100-200$ kpc (e.g., De Grandi et al.\ 2004; we note that 
this still remains the case when one uses higher spatial 
resolution {\it Chandra} data; see Vikhlinin et al.\ 2005).  
This is roughly a factor of 3-7 larger than the typical 
effective radius of the central BCG.  If the metals that make 
up the peak originated from the stellar component of the 
central BCG, then some mechanism is required to lift the metals 
to larger radii.  Mathews et al.\ (2004) have shown that 
central heating and convective outflows can provide the 
necessary transport mechanism to explain the 
width of the observed gradients.  We perform a slightly 
different test using our model.  In particular, just prior to 
the ``wood stove'' switching on we assume the metallicity 
distribution of our model 
clusters is identical to the average CC cluster profile observed 
by De Grandi et al.\ (see their Fig.\ 2).  We then track the 
evolution of this profile after the AGN heating turns on.  
We find that for the range of central heating strengths 
explored above, the metallicity gradients does not evolve 
significantly for many ($>5$) gigayears.  The reason for this 
behaviour is two-fold.  First, even though some material 
buoyantly rises to very large radii, a sizeable fraction of it 
remains within radii that are smaller than the width of 
the peak.  So the peak is maintained to some degree by the 
feedback loop.  Second, and more importantly, the width of 
the peak is comparable to the cooling radius.  This means it 
will take nearly a Hubble time for all the metals to enter into 
the feedback loop just once.  Furthermore, we have neglected 
additional metal injection due to stellar evolution of the 
central BCG and the effects of sedimentation.  Therefore, in 
accordance with Mathews et al.\ , we find that the observed 
metallicity gradients are extremely robust to strong central 
heating and its associated convective outflows.

\subsection*{E.2 Comparison to observed X-ray bubbles}

The ``wood stove'' heating model assumes that the ICM at small 
radii is reheated by some mechanism and then convects to larger 
radii perhaps in the form of ``bubbles'', such as those commonly 
observed in CC clusters.  In addition to the reheated ICM, the 
bubbles may also contain a relativistic component in the form 
of magnetic fields and/or relativistic particles.  Analysis of 
radio data appears to confirm this is the case for several 
well-known sets of observed bubbles (e.g., B{\^i}rzan et al.\ 
2004; Dunn et al.\ 2005).  In order for the preheating plus 
``wood stove'' heating model to be successful, virtually all 
of the cooling ICM at small radii must be reheated and entrained 
within the rising bubbles, otherwise cool gas will rapidly pile 
up at the cluster centre, violating observational constraints. 
Therefore, there is an upper limit to the fraction of bubble 
material that can reside in a relativistic component and still 
retain consistency with the (reheated ICM) mass transfer rate 
required for the ``wood stove'' heating model to be successful.  
It is therefore interesting to calculate whether the observed 
bubble sizes, duty cycles, and relativistic content are 
consistent with the proposed model.

We have tested the above as follows.  In our preheating + 
AGN ``wood stove'' model, the central mass inflow rate is 
$\dot{M} \sim 20 \ M_\odot$ yr$^{-1}$ for several gigayears 
following AGN ignition.  This is the rate at which material 
must be recycled in the rising bubbles.  The mass of each 
bubble is set by selecting a bubble duty cycle, $t_{\rm bub}$; 
i.e., $M_{\rm bub} = \dot{M} t_{\rm bub}$.  The density of 
the reheated ICM inside the bubbles can then be calculated if the 
bubble size is known; i.e., assuming spherical symmetry, 
$\rho_{\rm bub} = M_{\rm bub}/(4 \pi r_{\rm bub}^3 /3)$.  The 
pressure of the (non-relativistic) reheated ICM can be 
calculated 
as $P_{\rm bub, non-rel} \propto S_{\rm bub} \rho_{\rm 
bub}^{5/3}$, if the entropy of the reheated ICM, $S_{\rm bub}$, 
is known.  In addition to reheated ICM, we assume the rising 
bubbles contain a relativistic component with pressure $P_{\rm 
bub, rel}$.  Assuming the bubble is in pressure equilibrium with 
the ambient ICM, we therefore have $P_{\rm bub, non-rel} + P_{\rm 
bub, rel} = P_{\rm bub, tot} = P_{\rm ICM}$.  The relativistic 
content of the bubble, $P_{\rm bub, rel}/P_{\rm bub, tot}$, is 
therefore fixed when appropriate choices for $t_{\rm bub}$, 
$r_{\rm bub}$, and $S_{\rm bub}$ are made.

The bubble duty cycle, $t_{\rm bub}$, of observed clusters 
is difficult to pin down exactly (e.g., B{\^i}rzan et al.\ 
2004), but it can be said with some certainty that it typically 
exceeds $10^7$ yr, since most CC clusters with bubbles contain 
only a single set of (visible) bubbles and have inferred ages of 
roughly this long (e.g., Dunn et al.\ 2005).  If AGN bubble 
heating is triggered at the onset of catastrophic cooling, then 
we should expect $t_{\rm bub}$ to be of order $10^8$ years.  In 
addition, observed bubbles typically have radii ranging from 5 kpc $< 
r_{\rm bub} < 20$ kpc (B{\^i}rzan et al.\ 2004).  Unfortunately, 
there are currently no strong constraints on the properties of 
the reheated ICM contained within bubbles (if, in fact, bubbles 
contain reheated ICM).  In order for the bubbles to appear as 
surface brightness depressions in {\it Chandra} X-ray images, 
though, the reheated material must be fairly hot ($kT > 15$ 
keV).  Assuming a duty cycle of $10^8$ yrs and a bubble radius 
of 10 kpc, the reheated ICM must therefore have an entropy, $S_{\rm 
bub}$, greater than about 200 keV cm$^2$.  As we demonstrated 
above (in Fig.\ 12), convective outflows driven by this 
level of heating yield an excellent fit to the observed entropy 
profiles of CC clusters for a fairly broad range of preheating 
levels.  If the entropy of the reheated material is assumed to 
be at 
this minimum level of $200$ keV cm$^2$, we find the relativistic 
component contributes $\approx 30\%$ of the total bubble pressure.  
Increasing the entropy of the reheated ICM has the effect of 
lowering $P_{\rm bub, rel}$.  For example, we find the entropy of 
the reheated material cannot exceed much beyond $300$ keV cm$^2$, 
otherwise the model requires $P_{\rm bub, rel} \leq 0$.  However, 
it should be noted that our quoted results are specific to the 
$r_{\rm bub} = 10$ kpc case.  Adopting a larger bubble radius, 
say of 20 kpc, will lead to a more important contribution from 
the relativistic component.

Clearly better constraints on the properties of the reheated ICM 
and the sizes of bubbles are needed to pin down $P_{\rm bub, 
rel}/P_{\rm bub, tot}$ (or vice-versa).  However, the above 
demonstrates that the proposed model is at the very 
least consistent with the the presence of relativistic material 
in some bubbles.  This is encouraging since it certainly did not 
have to be the case.  

\subsection*{E.3 Energetics and distributed heating}

In the simple heating model explored in \S 5, all of the 
(post-preheating) energy is assumed to be deposited near the 
centre of the cluster.  This raises the entropy of the gas there 
and it begins to rise buoyantly, perhaps in the form similar to 
observed bubbles\footnote{Note the transported material could, 
in addition to reheated ICM, be composed of entrained 
cooler material which is also deposited at large radii and 
mixed with the ICM there.}.  However, conventional wisdom is 
that energy must be deposited at all radii to offset cooling 
and the heating process must be relatively gentle.  This raises 
several important questions about the energetics of the 
preheating + AGN ``wood stove'' model.  For example: How much 
energy does the model require in order to stave off catastrophic 
cooling at the cluster centre?  Will injecting this energy 
cause large entropy inversions in the ICM?  How can heating 
just the central regions solve the long term problem that all 
of the gas within the cooling radius will cool significantly 
over the lifetime of the cluster?  What about distributed 
heating?  Let us now consider these questions in order.

In the preheating + ``wood stove'' model, the central mass 
inflow rate (that is, after central heating is initiated) is 
typically $\dot{M} \sim 20 \ M_\odot$ yr$^{-1}$ for a 
typical $M_{200} = 10^{15} M_\odot$ cluster.  This is the 
minimum rate of mass that must be recycled in the flow.  
Adopting this rate and the central gas density of our model, we 
find the time-averaged rate of thermal energy injection, 
$<\dot{E}>$, is 

\begin{eqnarray}
<\dot{E}> \approx 4\times10^{44} \ {\rm ergs/s} \ \biggl(\frac{S}{200 
{\rm \ keV \ cm}^2} \biggr) \biggl(\frac{\dot{M}}{20 \ M_\odot 
\ {\rm yr}^{-1}} \biggr) \nonumber
\end{eqnarray}

\noindent where $S$ is the entropy of the reheated gas inside 
the bubble.  The energy required to operate the central 
feedback loop is therefore typically lower than most other 
heating models (that distribute heat throughout the volume 
enclosed by the cooling radius) and is within reach of observed 
AGN power output.

Will introducing all of this energy at the centre cause an 
entropy inversion?  This is an important constraint, since 
inversions in the ambient ICM are not typically observed in CC 
clusters.  First, we note that the proposed model differs from 
most other heating models in that it assumes that the observed 
bubbles are made up of the heated material (instead of the 
bubbles being comprised entirely of relativistic material).  
So, in this sense, {\it the observed bubbles are entropy 
inversions.}  Importantly, the reheated material must be 
hot enough such that the bubbles appear as X-ray surface 
brightness depressions that do not contribute much to the X-ray 
flux or the mean radial profiles derived for CC clusters (as 
observed).  As noted above, this requires an entropy of 
about $200$ keV cm$^2$ or higher.  Whether or not material {\it 
outside} the bubble is heated significantly by the inflation of 
the bubble to cause an entropy inversion in the ambient ICM 
depends on how fast the bubble is inflated (i.e., super- or 
subsonically; see, e.g., Nusser et al.\ 2006).  If the bubble 
is inflated gently, then no significant heating of  the ambient 
ICM need occur during the inflation of the bubble.  This would 
be consistent with the observed lack of strong shocks near most 
young bubbles (e.g., Fabian et al.\ 2000; Blanton et al.\ 
2001).

How does the ``wood stove'' model avoid the long term cooling 
problem?  
It is clear that since neither preheating nor adiabatically 
rising bubbles are able to balance radiative cooling losses at 
larger radii (but still within $r_{\rm cool}$), the model we 
have proposed cannot avoid the long term cooling problem 
indefinitely.  However, if the level of preheating is 
sufficiently high ($\sim 200-300$ keV cm$^2$ for $\sim 10^{15} 
M_\odot$ clusters), the long term cooling problem is severely 
mitigated.  The reason for this is that because the system 
starts off on a higher adiabat, the initial cooling time of the 
ICM at all radii is significantly increased relative to the 
no preheating case.  Consequently, clusters that have been 
significantly preheated will not undergo catastrophic at the 
centre until many gigayears following cluster formation.  At 
the onset of catastrophic cooling at the cluster centre, the 
AGN switches on and heats the coolest gas.  It is important to 
recognise, however, that in the preheating case the onset 
of catastrophic cooling at the very centre {\it does not} 
coincide with the time when the cluster entropy profile at larger 
asymptotes to the pure cooling state.  As one can see in Fig.\ 
12, the tendency of the model is indeed to evolve towards the 
pure cooling state, as expected.  But because preheating 
reduces the initial cooling rate of the gas at larger radii, 
the time required to complete this transition is long compared 
to the typical age of massive clusters ($\sim 5$ Gyr, according 
to cosmological simulations) or between the onset of 
catastrophic cooling at the centre and the present day.  In 
other words, when preheating is introduced the long term 
cooling problem is not really ``long term'' at all.  A clear 
prediction of the model, therefore, is that CC clusters should 
be much less abundant at high redshifts.  Recent observations 
appear to suggest that this is indeed the case (Vikhlinin et 
al.\ 2007; Santos et al.\ 2008).

If we continue to evolve the model into the future, a further 
prediction is that the CC clusters we observe today will, in 
several gigayears time, complete their transition to states 
similar to those plotted in Fig.\ 11 (i.e., the no preheating 
case).  By the same logic, many NCC clusters are expected to 
slowly evolve into systems that resemble present-day CC 
clusters.  In this respect, our model differs from many other 
recently proposed models that seek to balance the heating and 
cooling rates of clusters, and thus arrive at a perfect 
steady-state configuration for the ICM.  It is worth noting 
that there are several different pieces of observational 
evidence that support a scenario in which cooling and heating 
are not balanced and, as a result, the ICM is slowly evolving 
in a manner similar to that proposed in the present study.  
These include: (1) the lack of CC clusters at $z > 0.5$ 
reported recently by Vikhlinin et al.\ (2007); (2) the 
accumulation of some cold gas (e.g., Edge et al.\ 2002) and 
moderate star formation rates (e.g., Egami et al.\ 2006) in the 
BCGs of some  extreme CC clusters; and (3) the scaling between 
total radio-AGN energy output and cluster velocity dispersion 
is observed to be considerably shallower than the scaling of 
the radiative cooling rate with cluster velocity dispersion 
(Best et al.\ 2007).

Finally, in its simplest form, the preheating plus AGN ``wood 
stove'' model neglects distributed heating by the bubbles as 
they rise buoyantly through the ICM.  However, most existing 
theoretical models and simulations of bubbles (e.g., Quilis et 
al.\ 2001; Br{\"u}ggen \& Kaiser 2002; Churazov et al.\ 2002; 
Omma \& Binney 2004; Robinson et al.\ 2004; Omma et al.\ 2004; 
Dalla 
Vecchia et al.\ 2004; Roychowdhury et al.\ 2004; Voit \& 
Donahue 2005; Heinz et al.\ 2006) show that the bubbles do not 
just float up to the level which their contents are at home and 
then simply melt away.  Instead some of the energy associated 
with the bubbles is expected to be dissipated in the ambient ICM 
as they rise (e.g., through shocks, dissipation of $PdV$ work 
as the bubbles rise and expand, dissipation of turbulent 
motions generated by the rising bubbles; see McNamara \& 
Nulsen 2007).  Since the estimated energy associated with 
observed bubbles is fairly large (typically $\sim 10^{60}$ 
ergs), distributed bubble heating could be quite significant 
and may even be capable of balancing cooling losses.  But, as 
noted earlier, the difficulty in completely balancing cooling 
losses lies in distributing the heat throughout the extremely 
large volume contained within the cooling radius and over the 
appropriate timescales.  Many recent simulations (e.g., Soker \& 
Pizzolato 2005; Vernaleo \& Reynolds 2006; Gardini 2007; 
Pavlovski et al.\ 2007) show that buoyantly rising bubbles may 
not deposit the heat where and when it is needed.  (Highly 
efficient thermal transport mechanisms, such as conduction or 
large physical viscosities, can help to distribute the heat 
more evenly and over larger scales, but may lead to a variety of 
other problems; see, e.g., Appendix D.)  Nevertheless, the 
assumption of the proposed model that the bubbles rise 
adiabatically is clearly unrealistic and should be addressed in 
further explorations of the model.  This will require the use of 
full hydrodynamic simulations, since distributed bubble heating 
is inherently a 3-D problem.

Having said the above, we do not expect that incorporating a 
more realistic treatment of the bubble heating will change the 
general concept of the proposed model.  The mass transport in 
CC clusters {\it must} be quite efficient if this mechanism is 
to explain the observed widths of metallicity peaks (which are 
much more extended than the light of the central BCG) 
in such systems.  Indeed, some recent observations show 
strong evidence that bubbles are transporting metals to large 
radii (e.g., Simionescu et al.\ 2007).  In addition, recent 
observational studies of emission-line filaments in a few CC 
systems have revealed that the filaments are outflowing and the 
flow patterns are consistent with simulations of buoyantly 
rising bubbles (e.g., Hatch et al.\ 2006).  Finally, 
hydrodynamic simulations of jets (e.g., Omma et al.\ 2004; 
Dalla Vecchia et al.\ 2004; Brighenti \& Mathews 
2006; Heinz et al.\ 2006; Sternberg et al.\ 2007; Br{\"u}ggen et 
al.\ 2007) demonstrate that large quantities of gas are being 
transported to large radii and mixed with the gas there.  All of 
these arguments provide support that the mass transport 
mechanism that lies at the heart of the ``wood stove'' model 
is indeed appropriate in CC clusters.

Finally, while there are several reasons to recommend the 
preheating + AGN ``wood stove'' model, we recognise that this 
model is not conventionally invoked to resolve the cooling 
problem in clusters.  Here, we would simply like to stress that 
while we have adopted this simple heating model as a convenient 
foil, the central result of this paper is that all clusters 
require preheating to {\it establish} their structural 
properties in the first place.  This conclusion does not rest on 
the assumed form of present-day heating.

\bsp

\label{lastpage}

\end{document}